\pdfoutput=1
\documentclass[11pt,twoside,a4paper,cmspaper,final,collab]{cms-tdr}
\def\svnVersion{e4f2801}\def\svnDate{2024/09/14}\def\cmsCernNoTag{CERN-EP-2024-120}\def\cmsCernDate{\today}\def\cmsMessage{Published in the European Physical Journal C as \href{http://dx.doi.org/10.1140/epjc/s10052-024-13244-0}{\doi{10.1140/epjc/s10052-024-13244-0}.}}
\begin{document}\cmsNoteHeader{BPH-23-005}

\newcommand{\pipi}      {\PGpp\PGpm}
\newcommand{\Dzorb}     {\PDz\,(\PADz)\xspace}
\newcommand{\KSKS}      {\ensuremath{\PKzS\PKzS}\xspace}
\newcommand{\KSPP}      {\ensuremath{\PKzS\pipi}\xspace}
\newcommand{\KSPIPI}    {\ensuremath{\KSPP}\xspace}
\newcommand{\pp}        {\ensuremath{\Pp\Pp}\xspace}
\newcommand{\DzPip}     {\ensuremath{\PDz\PGpp}\xspace}
\newcommand{\DzbPim}    {\ensuremath{\PADz\PGpm}\xspace}
\newcommand{\DPi}       {\ensuremath{\PD\PGppm}\xspace}
\newcommand{\mKSKS}     {\ensuremath{m(\KSKS)}\xspace}
\newcommand{\mDPi}      {\ensuremath{m(\DPi)}\xspace}
\newcommand{\DzKSKS}    {\ensuremath{\PDz\to\KSKS}\xspace}
\newcommand{\DzbKSKS}   {\ensuremath{\PADz\to\KSKS}\xspace}
\newcommand{\DzKSPP}    {\ensuremath{\PDz\to\KSPP}\xspace}
\newcommand{\DzKK}      {\ensuremath{\PDz\to\PKp\PKm}\xspace}
\newcommand{\DzPP}      {\ensuremath{\PDz\to\pipi}\xspace}
\newcommand{\DspDzPip}  {\ensuremath{\PDstp\to\PDz\PGpp}\xspace}
\newcommand{\DsmDzbPim} {\ensuremath{\PDstm\to\PADz\PGpm}\xspace}
\newcommand{\DspmDPi}   {\ensuremath{\PDstpm\to\DPi}\xspace}
\newcommand{\DsKSKSPI}  {\ensuremath{\PDpms\to\KSKS\PGppm}\xspace}
\newcommand{\BuDspmPi}  {\ensuremath{\PBp\to\PDstpm (\to \DPi)  X}\xspace}
\newcommand{\BdDspmPi}  {\ensuremath{\PBz\to\PDstpm (\to \DPi)  X}\xspace}
\newcommand{\chisq}     {\ensuremath{\chi^2}\xspace}
\newcommand{\pvtx}      {\ensuremath{P_{\text{vtx}}}\xspace}
\newcommand{\ptDst}     {\ensuremath{\pt (\PDstpm)}\xspace}
\ifthenelse{\boolean{cms@external}}{\newcommand{\cp}{\ensuremath{C\!P}\xspace}}{\newcommand{\cp}{\ensuremath{CP}\xspace}}
\newcommand{\acp}       {\ensuremath{A_{\cp}}\xspace}
\newcommand{\acpraw}    {\ensuremath{A_{\cp}^{\text{raw}}}\xspace}
\newcommand{\acpdet}    {\ensuremath{A_{\cp}^{\text{det}}}\xspace}
\newcommand{\acppro}    {\ensuremath{A_{\cp}^{\text{pro}}}\xspace}
\newcommand{\dacp}      {\ensuremath{\Delta\acp}\xspace}
\newcommand{\acpksks}   {\ensuremath{\acp(\KSKS)}\xspace}
\newcommand{\acpkspp}   {\ensuremath{\acp(\KSPP)}\xspace}

\newlength\cmsTabSkip\setlength{\cmsTabSkip}{1ex}
\ifthenelse{\boolean{cms@external}}{\providecommand{\cmsLeft}{upper\xspace}}{\providecommand{\cmsLeft}{left\xspace}}
\ifthenelse{\boolean{cms@external}}{\providecommand{\cmsRight}{lower\xspace}}{\providecommand{\cmsRight}{right\xspace}} 

\cmsNoteHeader{BPH-23-005}
\title{Search for \texorpdfstring{\cp}{CP} violation in \texorpdfstring{\DzKSKS}{D0 to K0S K0S} decays in proton-proton collisions at \texorpdfstring{$\sqrt{s} = 13\TeV$}{$\sqrt{s} = 13\TeV$}}

\author*[cern]{A. Cern Person}

\date{\today}

\abstract{A search is reported for charge-parity \cp violation in \DzKSKS decays, using data collected in 
	proton-proton collisions at $\sqrt{s} = 13\TeV$ recorded by the CMS experiment in 2018.
	The analysis uses a dedicated data set that corresponds to an integrated luminosity of 41.6\fbinv, 
	which consists of about 10 billion events containing a pair of \PQb hadrons, nearly all of which decay to charm hadrons.
	The flavor of the neutral \PD meson is determined by the pion charge in the reconstructed decays 
	\DspDzPip and \DsmDzbPim. The \cp asymmetry in \DzKSKS is measured to be 
	$\acpksks = (6.2 \pm 3.0 \pm 0.2 \pm 0.8)\%$, where the three uncertainties represent the statistical uncertainty, the systematic
	uncertainty, and the uncertainty in the measurement of the \cp asymmetry in the \DzKSPP decay. This is the first \cp asymmetry 
	measurement by CMS in the charm sector as well as the first to utilize a fully hadronic
	final state.}

\hypersetup{%
	pdfauthor={CMS Collaboration},%
	pdftitle={Search for CP violation in D0 to K0S K0S decays in proton-proton collisions at 13 TeV},%
	pdfsubject={CMS},%
	pdfkeywords={CMS, CP violation, charm physics, flavor physics, experimental results}}

\maketitle 

\section{Introduction \label{sec:introduction}}

\begin{figure}[ht]
	\centering
        \includegraphics[width=0.48\textwidth]{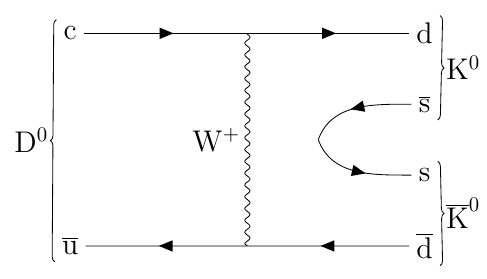}
        \includegraphics[width=0.48\textwidth]{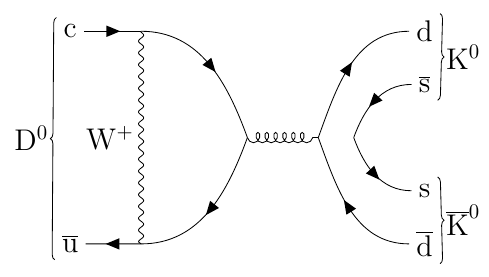}
	\caption{The decay of neutral charm meson to two neutral kaons: exchange (\cmsLeft) 
		and penguin annihilation (\cmsRight) diagrams. 
		\label{fig:diagrams}}
\end{figure}

{\tolerance=800
The noninvariance of fundamental interactions under the combined charge-parity (\cp)
transformation is one of the necessary conditions for the generation of the observed baryon asymmetry 
in the universe~\cite{AndreiDSakharov_1991}.
In the standard model (SM), the \cp symmetry violation originates from a single phase 
in the Cabibbo--Kobayashi--Maskawa (CKM) quark mixing matrix~\cite{PhysRevLett.10.531, Kobayashi1973}.
Extensive studies of \cp violation in weak interaction decays of strange and beauty mesons have been
performed by many experiments, with all results to date being consistent with the predictions based on the CKM formalism~\cite{PDG}.
However, the magnitude of \cp violation in the SM appears to be insufficient to explain the matter-antimatter 
asymmetry observed in the universe~\cite{43.120193.000331, 49.1.35, Hou:2008xd}, suggesting the existence of sources of \cp violation  beyond the SM. 
Charmed meson decays are the only meson decays involving an up-type quark where \cp violation can be studied, 
and are complementary to strange and beauty meson decays. 
In contrast to the \PK and \PB systems, \cp violation in charm mesons is severely suppressed by the 
Glashow--Iliopoulos--Maiani mechanism~\cite{Glashow:1970gm} and by the magnitude of the CKM elements~\cite{Kobayashi1973}. 
Given the strong SM suppression, an observation of a significant \cp violation in \PD meson 
decays may indicate a contribution from new physics, which can be different from those relevant for down-type quark systems.
The first observation of \cp violation in charm decays was recently reported by the LHCb Collaboration in a measurement of
the \cp asymmetry (\acp) difference between the \DzKK and \DzPP decays~\cite{LHCb:2019hro}. 
However, determining if this (or other) measurements
of \cp violation is an indication of new physics is hampered by large theoretical uncertainties associated with long-distance
contributions and nonperturbative effects~\cite{Lenz:2020awd}.

The \DzKSKS decay proceeds through the \PW boson exchange and penguin annihilation Feynman diagrams, 
some examples of which are shown in Fig.~\ref{fig:diagrams}, which results 
in a relatively small branching fraction of $( 1.41\pm 0.05) \times 10^{-4}$~\cite{PDG}. 
In this figure and throughout this paper, charge-conjugate states are implied, unless
otherwise indicated. Theoretical predictions indicate similar amplitudes and different phases for 
the two diagrams, which can result in \cp violation in this channel as large as a
few percent~\cite{PhysRevD.92.054036, PhysRevD.86.036012, PhysRevD.100.093002, PhysRevD.99.113001, PhysRevD.86.014023} 
and therefore possibly within reach of current experiments.
\par}

The \cp asymmetry \acp, for the \DzKSKS decay, is defined as
\begin{equation}\label{eq:acp}
    \acpksks = \frac{\Gamma(\DzKSKS)-\Gamma(\DzbKSKS)}{\Gamma(\DzKSKS)+\Gamma(\DzbKSKS)}.
\end{equation}
The current world average for the time-integrated \cp asymmetry is 
$\acpksks =(-1.9 \pm 1.1)\unit{\%}$~\cite{PDG}, which is dominated
by results from the LHCb~\cite{LHCb:2021rdn} and Belle~\cite{Dash:2017heu} Collaborations.

This paper presents the first \cp violation measurement by the CMS experiment in the charm sector. 
The flavor of the neutral \PD meson is determined from the pion charge found from reconstructing
the decays \DspDzPip and \DsmDzbPim. We measure the \cp asymmetry difference, \dacp, between the
signal channel \DzKSKS and the reference channel \DzKSPP. The \DzKSPP \cp asymmetry has been previously
measured~\cite{CDF:2012lpc} and found to be consistent with zero, as expected since this decay is not CKM-suppressed. 
Therefore, a significant deviation of \dacp from 0 would indicate \cp violation in the \DzKSKS decay.

In proton-proton $(\pp)$ collision data, the number of \PDstp 
and \PDstm decays (signal events, $N$) are measured, where both \PDz and \PADz are reconstructed in the 
\KSKS or \KSPP decay modes. The ``raw'' asymmetry between these numbers, 
\acpraw (defined in Eq.~\eqref{eq:acpraw}), is different 
from the true asymmetry \acp, due to the slightly
different production cross sections $(\sigma)$ of \PDstp and \PDstm mesons,
as well as to a possible difference in the detection efficiency ($\epsilon$) between 
\PDstp and \PDstm. Because these three
asymmetries are all small, the following relation can be written:
\begin{equation}\begin{aligned}
    \acp    \approx&\acpraw-\acppro-\acpdet,\quad\text{where}  \\
    \acpraw =&\frac{N(\DspDzPip)-N(\DsmDzbPim)}{N(\DspDzPip)+N(\DsmDzbPim)}, \label{eq:acpraw}\\
    \acppro =&\frac{\sigma_{\pp\to\PDstp X}-\sigma_{\pp\to\PDstm X}}{\sigma_{\pp\to\PDstp X}+\sigma_{\pp\to\PDstm X}},\quad\text{and} \\
    \acpdet =&\frac{\epsilon(\DspDzPip)-\epsilon(\DsmDzbPim)}{\epsilon(\DspDzPip)+\epsilon(\DsmDzbPim)}, 
\end{aligned}\end{equation}
where measuring the difference of \acp between the signal and reference channels, \acppro 
(\PDstpm production asymmetry) and \acpdet (\PDstpm detection asymmetry) 
cancel out, as they do not depend on the final state (\KSKS or \KSPP):
\begin{equation}\begin{aligned}
\ifthenelse{\boolean{cms@external}}
{ 
    \dacp &\equiv \acpksks - \acpkspp \\
    &= \acpraw(\KSKS) - \acpraw(\KSPIPI).
} 
{ 
    \dacp \equiv \acpksks - \acpkspp = \acpraw(\KSKS) - \acpraw(\KSPIPI).
} 
\end{aligned}\end{equation}
The reference channel was chosen to be as similar as possible in kinematics, topology, 
and final-state signature to the signal channel, to ensure that the 
reconstruction efficiency asymmetries cancel in the measured difference of asymmetries, \dacp.

{\tolerance=1200
The analysis uses proton-proton collisions data recorded by the CMS detector during 
the CERN LHC Run 2 in 2018, at $\sqrt{s}=13\TeV$. It utilizes the \PB parking data set~\cite{CMS:2024syx, CMS:2024zhe}, collected with
a set of single-muon triggers with different minimum thresholds on the muon transverse momentum (\pt) and impact parameter
with respect to the beamline. Different triggers were enabled depending on the instantaneous luminosity: as the luminosity decreased,
less restrictive triggers were enabled, as allowed by the limited event rate to be processed 
by the data acquisition system and recorded on tape. The data set contains about $1.2\times10^{10}$ events 
and corresponds to an integrated luminosity of $41.6\fbinv$. 
More details about this data set can be found in Ref.~\cite{CMS:2024syx}.
These triggers are intended to select events containing a semimuonic decay of a \PQb hadron (or a
semimuonic decay of \PQc hadron that originated from a \PQb-hadron decay).
Since the trigger requires muons inconsistent with being produced in the primary interaction, 
most of such muons come from semileptonic decays of beauty hadrons, hence approximately 80\% of the events in this sample include
\PQb hadrons~\cite{CMS:2024syx, CMS:2024zhe}. 
As beauty hadrons nearly always decay into charm hadrons, this data set also provides a rich sample of charm decays, 
making it suitable for \cp violation studies in the charm sector.
\par}

\section{The CMS detector \label{sec:cmsdet}}

The central feature of the CMS apparatus is a superconducting solenoid of 6\unit{m} internal dia\-meter, 
providing a magnetic field of 3.8\unit{T}. Within the solenoid volume are a silicon pixel and 
strip tracker, a lead tungstate crystal electromagnetic calorimeter, and a brass 
and scintillator hadron calorimeter, each composed of a barrel and two endcap sections. 
Forward calorimeters extend the pseudorapidity ($\eta$) coverage provided by the barrel and endcap detectors. 
Muons are measured in gas-ionization detectors embedded in the steel flux-return yoke outside the solenoid.
The reconstructed decays used by this analysis contain five pions in the final state. Pions are measured by the silicon tracker whose
setup during the 2018 LHC running period, when the data used in this paper were recorded, 
consisted of 1856 silicon pixel~\cite{Phase1Pixel} and 15\,148 silicon strip detector modules. 
For non-isolated particles with $\abs{\eta} < 3$ and $1 < \pt < 10\GeV$, the track resolutions are typically 1.5\% in \pt and 
20--75\mum in the transverse impact parameter~\cite{DP-2020-049}.

Events of interest are selected using a two-tiered trigger system. The first level, 
composed of custom hardware processors, uses information from the calorimeters and muon detectors 
to select events at a rate of around 100\unit{kHz} within a fixed latency of 
4\mus~\cite{CMS:2020cmk}. The second level, known as the high-level trigger, 
consists of a farm of processors running a version of the full event reconstruction software 
optimized for fast processing, and further reduces the event rate before data storage~\cite{CMS:2016ngn}. 

A more detailed description of the CMS detector, together with a definition of the coordinate system used 
and the relevant kinematic variables, can be found in Ref.~\cite{CMS:2008xjf}. 

\section{Simulated event samples \label{sec:simevts}}

The simulated event samples used in this analysis are generated with \PYTHIA 8.230~\cite{SJOSTRAND2015159}.
The \PYTHIA output is interfaced with \EVTGEN~\cite{Lange:2001uf} 1.3.0, which simulates various \PQb and \PQc hadron decays. 
The underlying event is also modeled with \PYTHIA using the CP5~\cite{CMS:2019csb} tune. Final-state photon 
radiation is modeled with \PHOTOS 3.61~\cite{Barberio:1993qi}. Samples with inclusive decays \BuDspmPi, \BdDspmPi, and prompt \DspmDPi were generated.
The events were then passed through a detailed \GEANTfour-based 
simulation~\cite{AGOSTINELLI2003250} of the CMS detector, followed by the trigger and reconstruction 
algorithms identical to those used for the collision data. 

\section{Reconstruction of charm meson decays\label{sec:evesel}}

The reconstruction starts with finding $\PKzS\to\pipi$ candidates as described in Ref.~\cite{CMS_tracks_V0}.
The two oppositely-charged pion tracks are fit to a common vertex that is required to have a \chisq fit probability $\pvtx > 1\%$. 
The dipion invariant mass must be within 20\MeV of the world average value of the \PKzS meson mass~\cite{PDG}, 
corresponding to approximately three times the mass resolution. 

In the signal channel, two $\PKzS\to\pipi$ candidates are each fit again with kinematic constraints to the \PKzS mass, and subsequently, 
the \PKzS candidates are fitted as two virtual tracks to a common vertex, assumed to be the \DzKSKS decay vertex. 
The \KSKS invariant mass is required to be between 1.77 and 1.95\gev, and the vertex fit probability 
must exceed 1\%. Both \PKzS decay vertices have to be displaced 
in three-dimensional (3D) space by at least one standard deviation (s.d.) from the fitted \KSKS vertex, 
and the corresponding pointing angle (the angle between the particle
momentum and the vector joining the production vertex with its decay vertex) for each \PKzS candidate
is required to be less than $90^\circ$. 

In the reference channel, after the single \PKzS selection, two additional high-purity~\cite{Chatrchyan:2014fea} and opposite-sign
tracks with $\pt>0.6\gev$ (and at least one of them with $\pt>0.7\gev$) are selected.
The \KSPP combination is then fit to a common vertex, assumed to be the \PDz decay vertex, 
which must have a $\pvtx > 5\%$, and an invariant mass between 
1.823 and 1.908\gev, assuming charged-pion mass~\cite{PDG} for both tracks, which corresponds to approximately twice the mass resolution.

The primary vertex (PV) is selected from the reconstructed \pp interaction vertices as the one 
with the smallest pointing angle of the \PDz candidate. 
After the \PDz reconstruction, an additional track is 
added to form the \DspDzPip or \DsmDzbPim candidates. A two-object vertex fit is 
performed to reconstruct the \PDstpm decay vertex, which is required to have a $\pvtx > 1\%$.
The \PDstp candidate invariant mass is determined from the refitted pion and \PDz four-momenta  and then corrected by subtracting the difference
between the reconstructed \PDz candidate mass and the world-average \PDz mass, to remove the effect of the \PDz detector mass resolution.  
The candidates are rejected if they are compatible with an incorrect decay topology that assumes negligible decay 
time of any of the \PKzS candidates.

\section{Final selection criteria \label{sec:optimiz}}

A mixture of different triggers with varying thresholds in the data set makes it challenging to 
properly model the kinematic distributions of charm mesons in the simulation. Therefore, 
an optimization of the selection criteria is done using the experimental data directly. 
A two-dimensional (2D) fit to the distribution of \mDPi vs \mKSKS, similar to the one described below, is 
performed for the data (with the \PDstp and \PDstm samples merged) while the selection criteria are varied.
The variables used in the optimization include the candidate \pt, $\eta$, 
\pvtx, distances between production and decay vertices for \PKzS and \PDz candidates divided by their corresponding uncertainties,
and corresponding pointing angles. The optimal criteria were chosen as those which result in the smallest relative uncertainty 
on the fitted signal yield. Cross-validation was used to ensure there is no bias due to statistical 
fluctuations in the data, via randomly splitting the data into six equal sub-samples, finding optimal criteria using five
of them and applying them to the last part. The procedure is repeated six times (each time leaving out a different 
part of the full data set) and results in six almost identical sets of selection criteria. 
The average value for each selection is taken as the final selection criteria, presented in Table~\ref{tab:cuts}. 

\begin{table*}[hbt]
\centering
\topcaption{
    Optimized selection criteria in the signal channel \KSKS. The requirements 
    on the \PKzS candidates in the third and fourth lines are given first for the \PKzS 
    with larger \pt, then for the \PKzS with lower \pt. \label{tab:cuts}}
\renewcommand{\arraystretch}{1.1}
\begin{tabular}{l l} 
    Variable                                                            &   Requirement             \\  
    \hline
    $\abs{\eta}$ of the tagging pion from \DspmDPi                      & $<$1.2                    \\
    \pt of the tagging pion from \DspmDPi                               & $>$0.35\gev               \\
    $\pt(\PKzS)$                                                        & $>$2.2 and $>$1.0\gev     \\
   \PKzS vertex displacement significance from the \PDz vertex in $xyz$ & $>$7 and $>$9             \\
    \PDz vertex displacement significance from the PV in $xy$           & $>$2                      \\
    \PDz vertex displacement significance from the PV in $xyz$          & $>$9        	            \\
    $\pvtx(\DPi)$                                                       & $>$5\%                    \\
    $\pvtx(\KSKS)$                                                      & $>$1\%                    \\
    $\pvtx(\pipi)$ for $\PKzS\to\pipi$                                  & $>$1\%                    \\
    Angle between \PDz momentum and displacement from PV in $xyz$       & $<$0.205\unit{rad}        \\
    Angle between \PDz momentum and displacement from PV in $xy$        & $<$0.237\unit{rad}        \\
    Angle between \PDz momentum and displacement from beamline in $xy$  & $<$0.237\unit{rad} 
\end{tabular}
\end{table*}

Similar selection criteria are applied to the reference channel, to minimize the differences in 
kinematic distributions between the signal and the reference channels: the only adjustment is that the scalar
sum of the \pt of the two pions that are not from the \PKzS decay in the reference channel must exceed 1\GeV and the single
\PKzS candidate in the reference channel must satisfy the requirements applied to the high-\pt \PKzS candidate in the signal channel.

\section{\texorpdfstring{\acp}{ACP} measurement: reference channel\label{sec:sigN}} 

The signal and reference channels are found to have consistent $\eta$ and $\phi$ distributions, but 
slightly different \ptDst ones, and thus the detection and production asymmetries 
may not cancel out fully in the \dacp measurement.
In order to suppress this effect, the reference channel data are reweighted to match 
the \ptDst distribution found in the signal channel, before splitting the samples by the pion charge.

\begin{figure*}[hbt]
	\centering
    \includegraphics[width=0.48\textwidth]{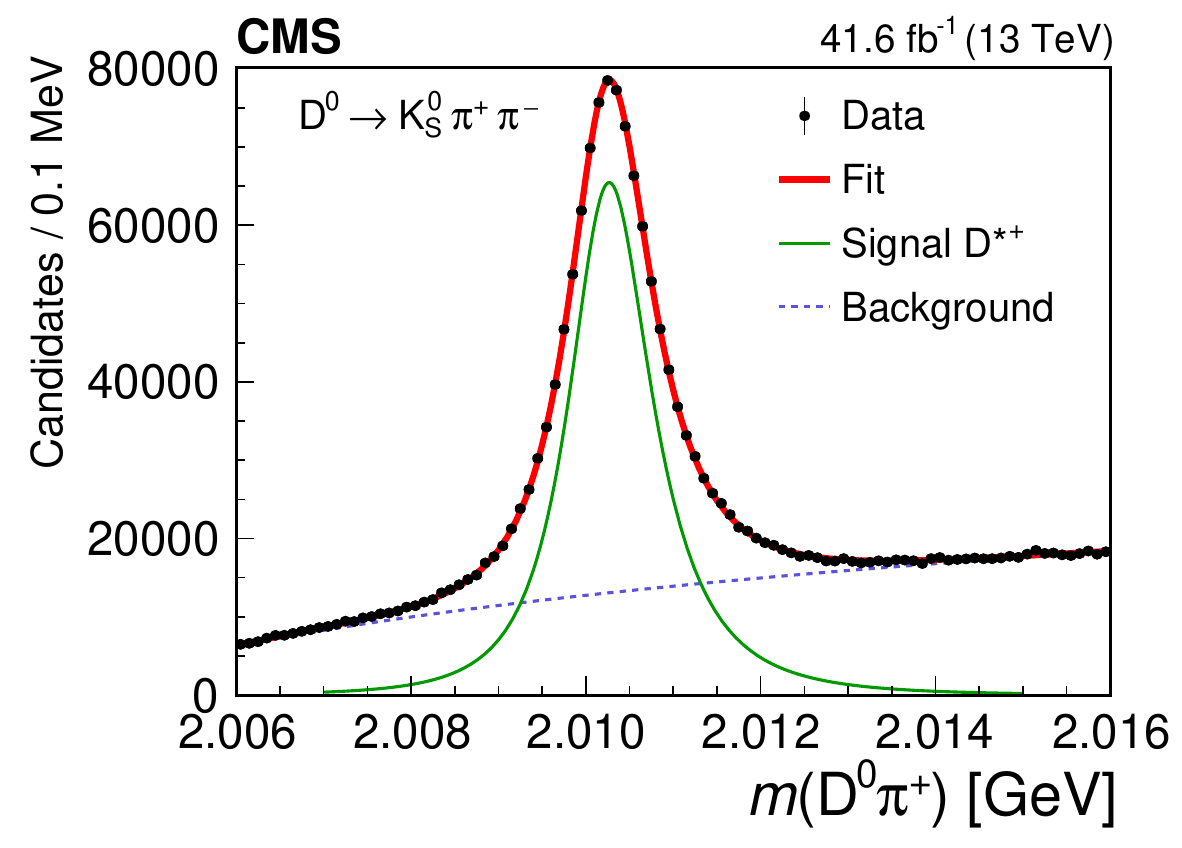}
    \includegraphics[width=0.48\textwidth]{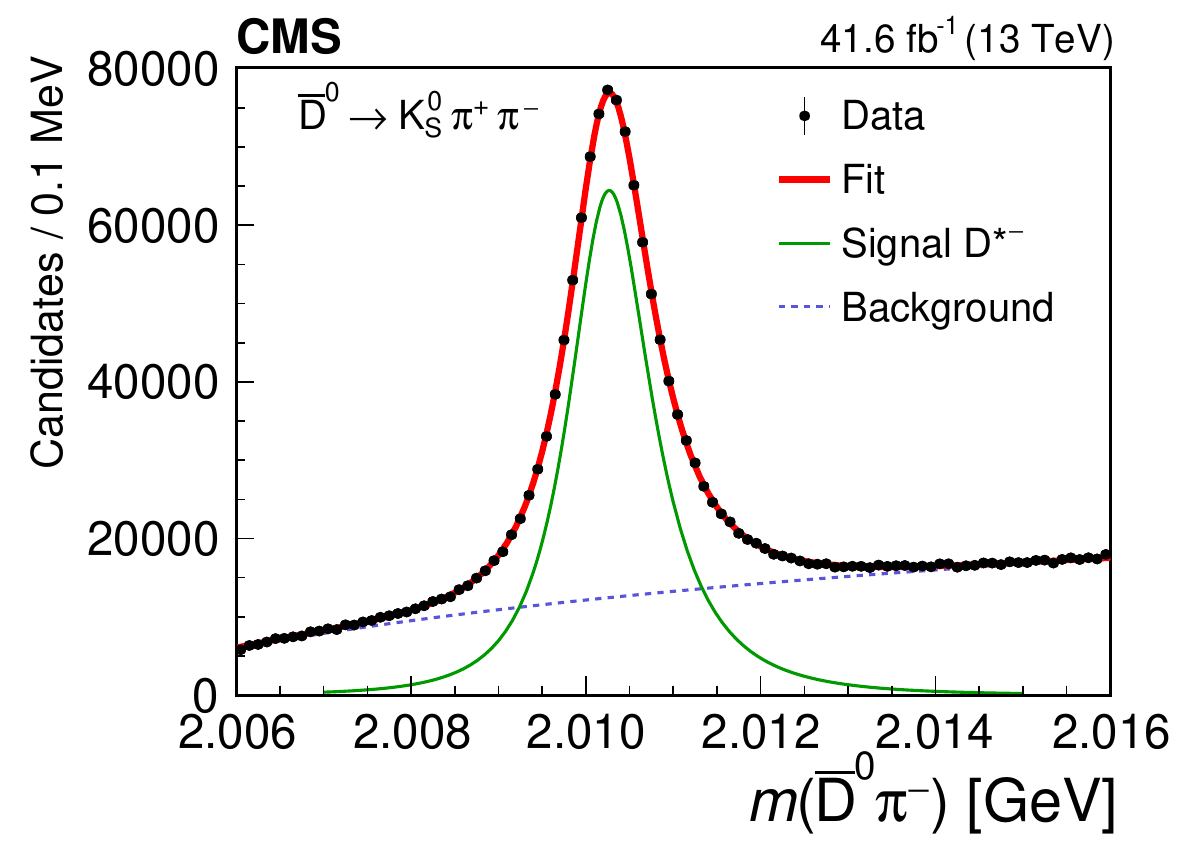}
	\caption{The \DzPip (left) and \DzbPim (right) invariant mass distributions for 
		     the \KSPP channel, with the result of the fit to both distributions. \label{fig:Dstnorm1}}
\end{figure*}

To extract the raw \cp asymmetry, a simultaneous binned extended maximum likelihood fit is performed 
on the invariant mass distributions \mDPi of weighted \PDstp and \PDstm candidates. 
The signal in $x = \mDPi$ is fitted with the $S_U$ Johnson transformation of the normal distribution~\cite{Johnson}, 
with the shape parameters shared between the \PDstp and \PDstm components while the signal yields are independent.
The background is modeled with a modified threshold function 
$(x-x_0)^\alpha (1 + a x)$, where $x_0$ is the threshold value equal to 
the sum of the masses of \PDz and \PGpp, and $\alpha$ and $a$ are floated in the fit and they are not shared
between the \PDstp and the \PDstm background model.
The results of the fit to the \mDPi distributions 
are presented in Fig.~\ref{fig:Dstnorm1} and Table~\ref{tab:Dstnorm1}.
The measured raw asymmetry is $\acpraw(\KSPP)= (0.78\pm 0.10)\%$, where the uncertainty is statistical only 
and accounts for the correlations found in the simultaneous fit.

\begin{table}[ht]
	\centering
	\topcaption{
		Results of the fit to the selected \DspDzPip and \DsmDzbPim candidates, where $\Dzorb\to\KSPIPI$.
		The \PDstpm signal yields $N$ given in the second column are used in the evaluation of \acpraw.
		The uncertainties are statistical only. \label{tab:Dstnorm1}}
	\renewcommand{\arraystretch}{1.1}
	\begin{tabular}{l c c} 
		Decay       & $N$                   & \chisq with 100 bins  \\ \hline
		\DspDzPip   & $944\,800\pm 3\,500$  &   78                  \\ 
		\DsmDzbPim  & $930\,150\pm 3\,400$  &   93				 
	\end{tabular}
\end{table}

\section{\texorpdfstring{\acp}{ACP} measurement: signal channel \label{sec:sigP}}
{\tolerance=800
To reduce the statistical uncertainty arising from the signal channel yield, which
is the dominant uncertainty in the analysis, the signal extraction is performed using a 2D unbinned
maximum likelihood fit performed simultaneously on the \PDstp and \PDstm samples 
to the distribution of \mDPi \vs \mKSKS. 
The fit function consists of the following components:
\begin{itemize}
	\item $\PDz \,\times\,\PDstp$, the signal component;
	\item $\PDz \,\times\,\, bkg$, for events containing genuine \PDz and background pion combinations;
	\item $bkg\,\,\times\,\,bkg$, for the background in both dimensions,
\end{itemize}
where each component is a product of two one-dimensional (1D) functions.
For the \PDstpm signal, the Johnson function is used with all signal shape parameters fixed to those 
found in the fit to the reference channel. This approach is verified to be reasonable using simulated event samples. 
The \PDz signal is modeled with a sum of two Johnson functions, 
all parameters of which are fixed to values determined from the simulation, except for a single free parameter that is 
used to scale the width.  
The background in $x = \mDPi$ is modeled with the same function as in the reference channel. 
The background in $y = \mKSKS$ is described with an exponential function $\exp(\beta y)$, 
where $\beta$ is floating in the fit, plus a
Gaussian function with free parameters to describe the partially-reconstructed background from the 
\DsKSKSPI decay producing an excess at about 1.83\gev in the \KSKS invariant mass distribution. 
\par}

{\tolerance=800
The projections of the data and the 2D fit on both axes are shown in Fig.~\ref{fig:TwoDfitP}; 
additional projections in sub-ranges are shown in Appendix~\ref{app}.
The fit results are listed in Table~\ref{tab:TwoDfitP}. The measured raw asymmetry is $\acpraw(\KSKS)= (7.1 \pm 3.0)\%$ and in combination 
with the results of Section~\ref{sec:sigN} the \acp difference is measured to be 
$\dacp= (6.3 \pm 3.0)\%$, where the uncertainty is statistical only 
and accounts for the correlations found in the simultaneous fit.
\par}
\begin{figure*}[htb]
    \centering
    \includegraphics[width=0.48\textwidth]{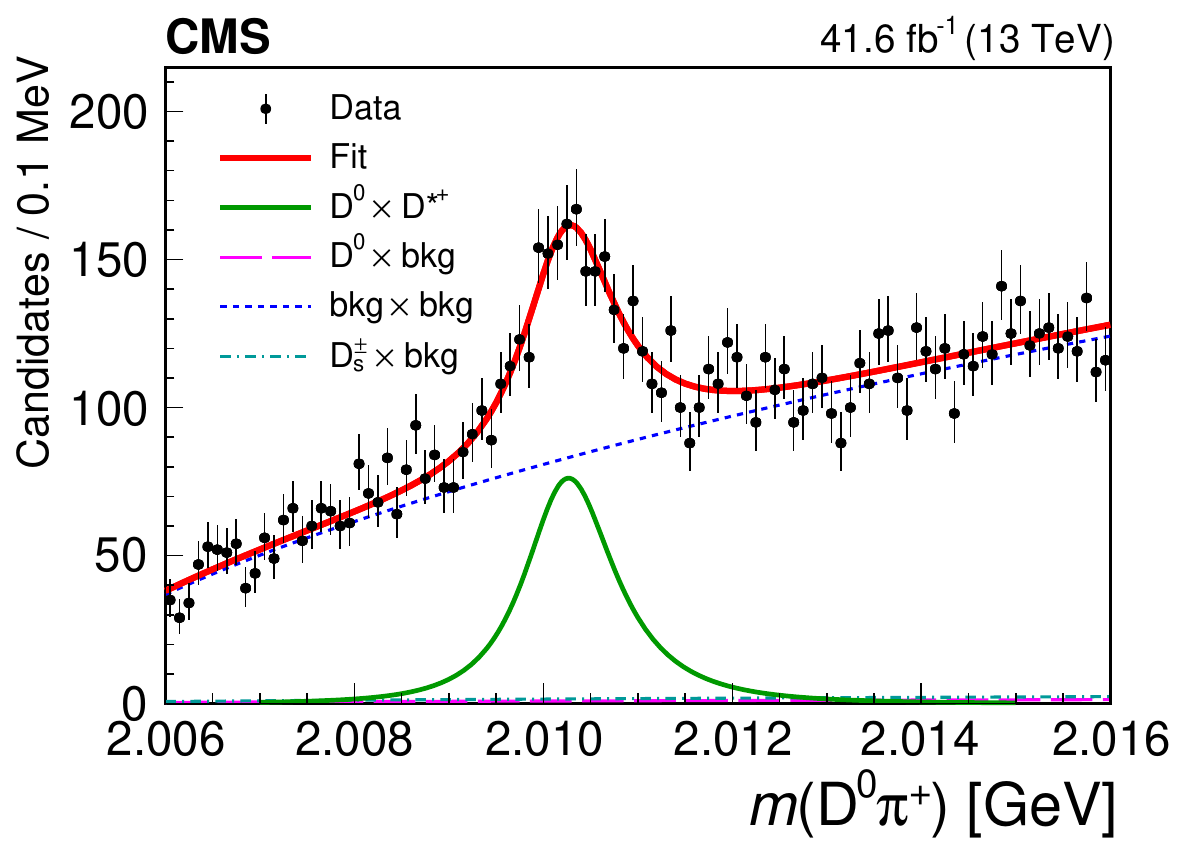}
    \includegraphics[width=0.48\textwidth]{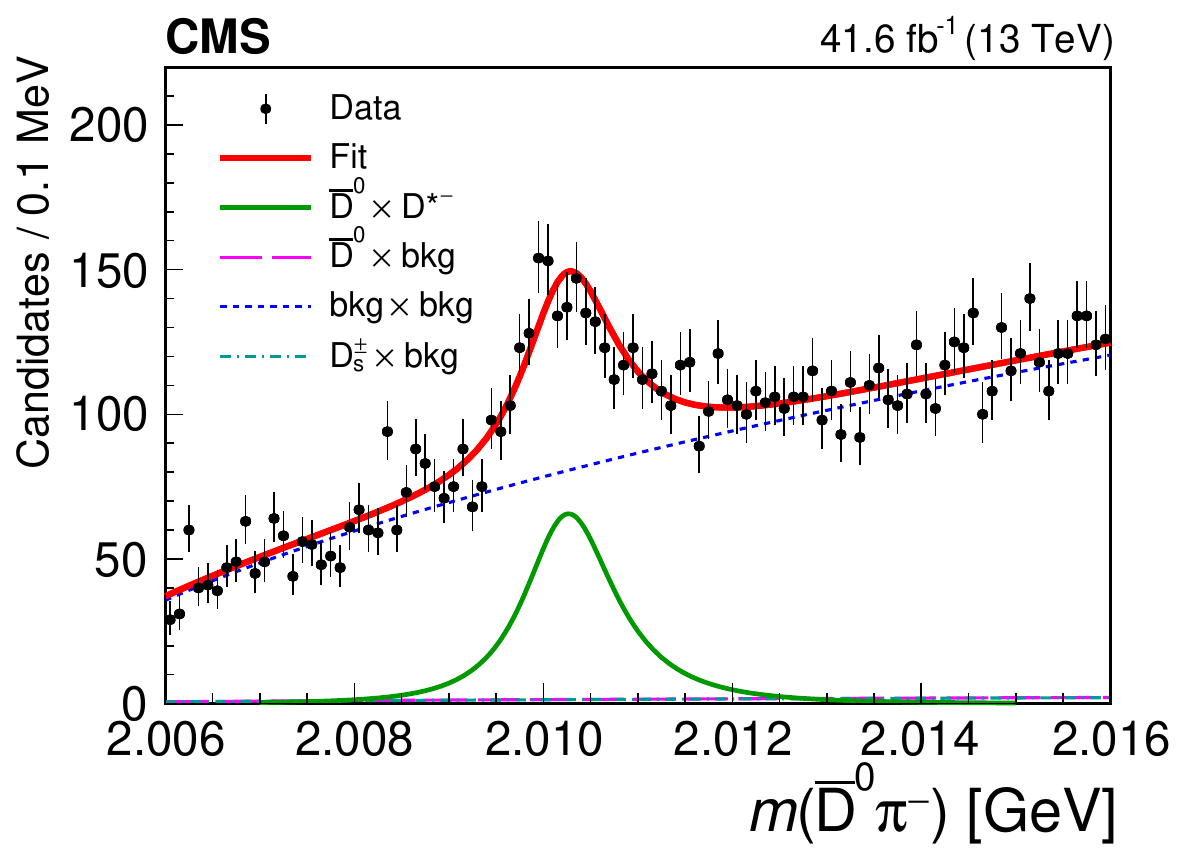}
    \includegraphics[width=0.48\textwidth]{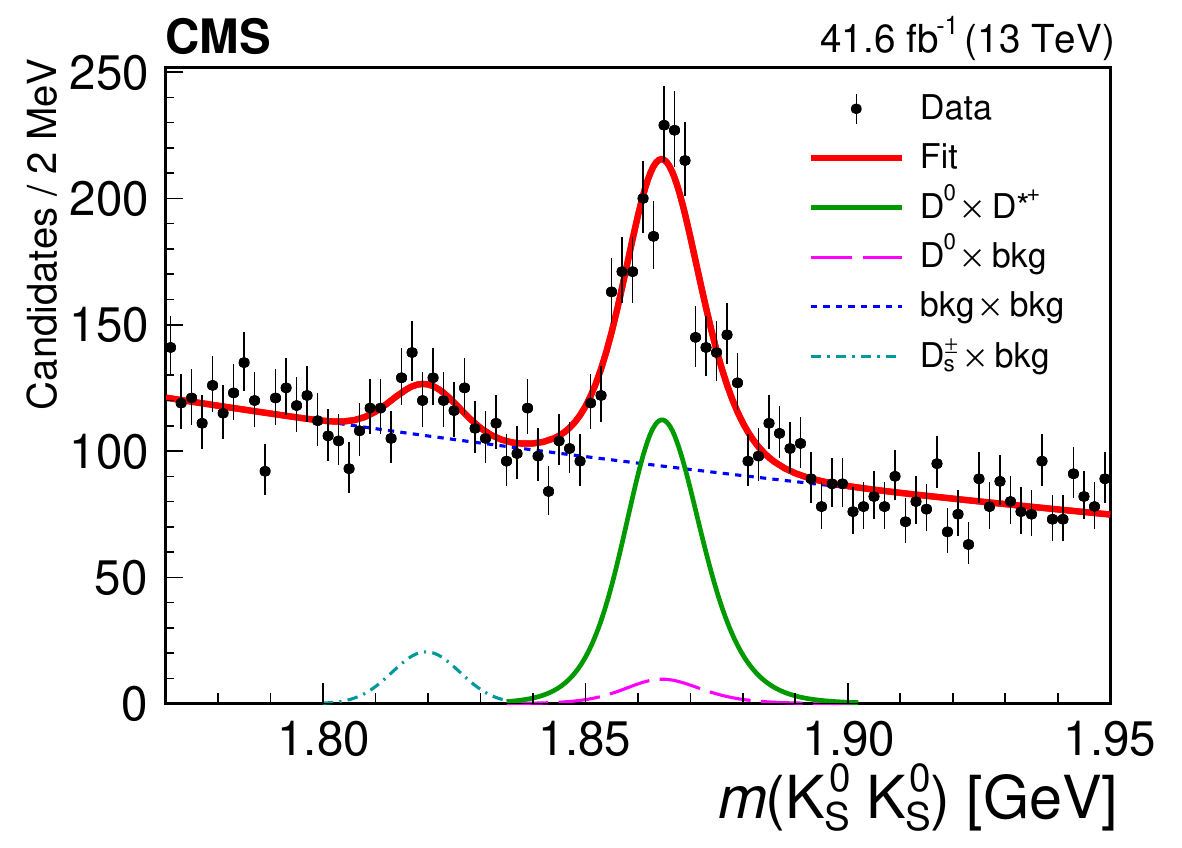}
    \includegraphics[width=0.48\textwidth]{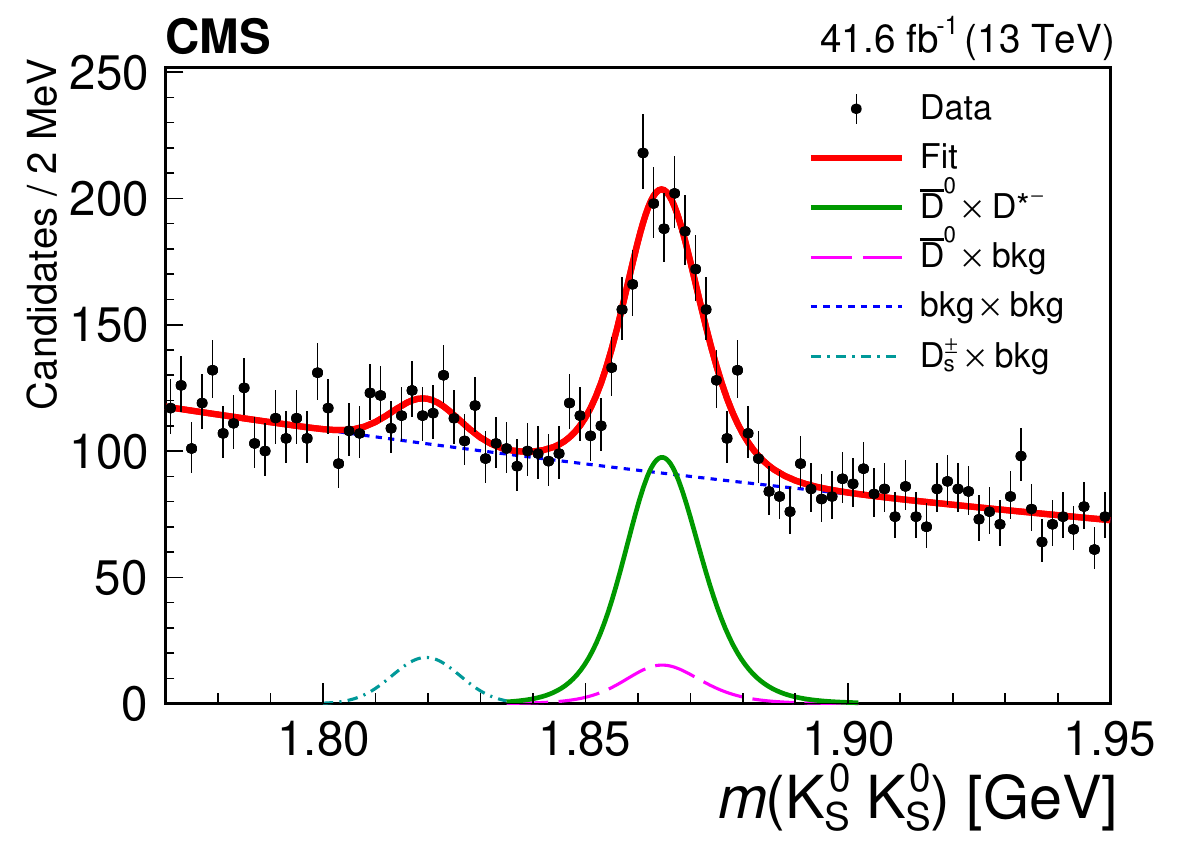}
    \caption{ The invariant mass distributions for \PDstp candidates (left) and \PDstm candidates (right),
    	 with the \mDPi distributions in the upper row and the \mKSKS distributions in
    	 the lower row. Projections of the simultaneous 2D fit are also shown.
        \label{fig:TwoDfitP}}
\end{figure*}

\begin{table}[ht]
    \centering
    \topcaption{
        Results of the 2D fit to the selected \DspDzPip and \DsmDzbPim candidates, where $\Dzorb\to\KSKS$.
        The \PDstpm signal yields $N$ given in the second column are used in the evaluation of \acpraw. The \chisq corresponds to the fit projection 
        with 100 bins in the $x = \mDPi$ axis and 90 bins in the $y = \mKSKS$ axis, as shown in Fig~\ref{fig:TwoDfitP}.
        The uncertainties are statistical only. \label{tab:TwoDfitP}}
    \renewcommand{\arraystretch}{1.1}
    \begin{tabular}{l c c c} 
        Decay       & $N$                   & \chisq ($x$ axis) & \chisq ($y$ axis) \\ \hline
        \DspDzPip   & $1 095 \pm 46$        &   77              & 90                \\ 
        \DsmDzbPim  & $\phantom{0}951\pm44$ &   93              & 62                 
    \end{tabular}
\end{table}

\section{Systematic uncertainties\label{sec:syst}}
The measured difference in the asymmetries is largely insensitive 
to many systematic uncertainties that would affect a measurement of 
\acp in a single channel, such as the difficult-to-measure production  
and detection asymmetries that would need a dedicated calibration procedure.

Uncertainties related to the choice of the signal and background models
are calculated separately using alternative models and assessing 
the observed variations in \dacp.

In the baseline approach, the signal in the \mDPi invariant mass distribution is modeled with the Johnson function~\cite{Johnson}.  
As an alternative, we use a Johnson+Gaussian function with a common mean. 
Another alternative is a sum of two Crystal Ball functions~\cite{CrystalBallRef}.
For each case, the reference channel is fit as a first step, then the 
obtained shape parameters are fixed in the 2D fit to the signal channel. 
Other components of the 2D fit remain unchanged from the baseline fit. 
The largest deviation in \dacp from the baseline value is taken as
a systematic uncertainty.

The baseline signal function for the \mKSKS invariant mass distribution
is a sum of two Johnson functions. As an alternative, we use a Johnson+Gaussian function
or a sum of two Crystal Ball functions. 
These variations have no effect on the fit of the reference channel, just on that of the signal channel. 
The largest deviation in \dacp from the baseline value is taken as
a systematic uncertainty.

The baseline background model in the $x = \mDPi$ distribution is 
$(x-x_0)^\alpha(1 + a x)$. 
An alternative background model is obtained
by changing the function multiplying the threshold function from
a linear polynomial to an exponential function. 
The baseline background model in the \mKSKS distribution is 
an exponential function and an exponential multiplied by a linear polynomial is used
as an alternative. These variations are taken as independent systematic uncertainties.

In the signal channel fit, there is a contribution from the \DsKSKSPI decay, 
which is modeled by a Gaussian with free parameters. 
As an alternative, we
remove this reflection by restricting the fit range to be $\mKSKS>1.835\gev$, and the deviation from the baseline 
is included as a systematic uncertainty.

To assess the systematic uncertainty related to the \pt reweighting, we 
vary the parameters of the reweighting function within their uncertainties.  
As an alternative, we consider the weights depending on the 
\pt of the low-momentum pion that is used for the flavor tagging instead of \ptDst.
The largest change is taken as a systematic uncertainty related to the reweighting.

Differences in \acppro and \acpdet between the two channels are expected to be reproduced by the 
simulation of the processes and the detector. Checking the reweighted reference channel and signal channel in simulation show
that the \pt-, $\eta$-, and $\phi$-dependent asymmetries are consistent with zero as is the integrated value 
of $(-0.13\pm 0.34)\%$. Therefore, no systematic uncertainty is assessed.

If multiple candidates in the same event are removed by keeping only the 
one with the highest \PDstpm vertex fit probability, the resulting 
\dacp changes negligibly and no corresponding systematic uncertainty is assigned.
Pion charge misidentification was shown to have a negligible effect as well.

All systematic uncertainties described above are uncorrelated and summarized in 
Table~\ref{tab:syst} together with the total systematic uncertainty, 
calculated as the sum in quadrature of the different contributions. 

\begin{table}[bth]
	\centering
	\topcaption{
		Absolute systematic uncertainties in the measurement of \dacp.
		\label{tab:syst}}
	\renewcommand{\arraystretch}{1.1}
	\begin{tabular}{lc}
		Source                     & Uncertainty (\%)   \\ \hline
		\mDPi signal model         & 0.10   \\
		\mDPi background model     & 0.02   \\
		\mKSKS signal model        & 0.04   \\
		\mKSKS background model    & 0.02   \\
		\mKSKS fit range           & 0.06   \\
		Reweighting 			   & 0.09   \\ [\cmsTabSkip]
		Total                      & 0.16
	\end{tabular}
\end{table}

\section{Summary\label{sec:results}}

A measurement of \cp violation in \PDz decays is reported,
using proton-proton collision data collected at $\sqrt{s} = 13\TeV$ with a novel high-rate data stream (\PB parking). 
These data correspond to an integrated luminosity of 41.6\fbinv and include about 10 billion events 
containing beauty hadron decays. The difference in the \cp asymmetries between 
\DzKSKS and \DzKSPP is measured to be:
\begin{equation}\begin{aligned}
\ifthenelse{\boolean{cms@external}}
{ 
    \dacp &\equiv \acpksks - \acpkspp = \\
    &= \left(6.3 \pm 3.0\stat \pm 0.2\syst\right) \%.
} 
{ 
    \dacp \equiv \acpksks - \acpkspp = \left(6.3 \pm 3.0\stat \pm 0.2\syst\right) \%.
} 
\end{aligned}\end{equation}

Using the world-average value of $\acpkspp = (-0.1 \pm 0.8)\%$~\cite{CDF:2012lpc,CLEO:2003sah,PDG}, 
we report the measurement
\begin{equation}
		\acpksks = (6.2 \pm 3.0 \pm 0.2 \pm 0.8)\%,
\end{equation}
where the three uncertainties represent the statistical uncertainty, the systematic
uncertainty, and the uncertainty in the measurement of the \cp asymmetry in the \DzKSPP decay.
The measured value is consistent with no \cp violation within 2.0 standard deviations. 
Likewise, it is consistent with the LHCb~\cite{LHCb:2021rdn} and the Belle measurements~\cite{Dash:2017heu}
at the level of 2.7 and 1.8 standard deviations, respectively.
Tabulated results are provided in the HEPData record for this analysis~\cite{hepdata}.
This is the first CMS search for \cp violation in the charm sector, paving the way 
for future measurements with more data, using new techniques, and in other channels.

\begin{acknowledgments}
We congratulate our colleagues in the CERN accelerator departments for the excellent performance of the LHC and thank the technical and administrative staffs at CERN and at other CMS institutes for their contributions to the success of the CMS effort. In addition, we gratefully acknowledge the computing centers and personnel of the Worldwide LHC Computing Grid and other centers for delivering so effectively the computing infrastructure essential to our analyses. Finally, we acknowledge the enduring support for the construction and operation of the LHC, the CMS detector, and the supporting computing infrastructure provided by the following funding agencies: SC (Armenia), BMBWF and FWF (Austria); FNRS and FWO (Belgium); CNPq, CAPES, FAPERJ, FAPERGS, and FAPESP (Brazil); MES and BNSF (Bulgaria); CERN; CAS, MoST, and NSFC (China); MINCIENCIAS (Colombia); MSES and CSF (Croatia); RIF (Cyprus); SENESCYT (Ecuador); ERC PRG, RVTT3 and MoER TK202 (Estonia); Academy of Finland, MEC, and HIP (Finland); CEA and CNRS/IN2P3 (France); SRNSF (Georgia); BMBF, DFG, and HGF (Germany); GSRI (Greece); NKFIH (Hungary); DAE and DST (India); IPM (Iran); SFI (Ireland); INFN (Italy); MSIP and NRF (Republic of Korea); MES (Latvia); LMTLT (Lithuania); MOE and UM (Malaysia); BUAP, CINVESTAV, CONACYT, LNS, SEP, and UASLP-FAI (Mexico); MOS (Montenegro); MBIE (New Zealand); PAEC (Pakistan); MES and NSC (Poland); FCT (Portugal); MESTD (Serbia); MCIN/AEI and PCTI (Spain); MOSTR (Sri Lanka); Swiss Funding Agencies (Switzerland); MST (Taipei); MHESI and NSTDA (Thailand); TUBITAK and TENMAK (Turkey); NASU (Ukraine); STFC (United Kingdom); DOE and NSF (USA).
	
\hyphenation{Rachada-pisek} Individuals have received support from the Marie-Curie program and the European Research Council and Horizon 2020 Grant, contract Nos.\ 675440, 724704, 752730, 758316, 765710, 824093, 101115353, 101002207, and COST Action CA16108 (European Union); the Leventis Foundation; the Alfred P.\ Sloan Foundation; the Alexander von Humboldt Foundation; the Science Committee, project no. 22rl-037 (Armenia); the Belgian Federal Science Policy Office; the Fonds pour la Formation \`a la Recherche dans l'Industrie et dans l'Agriculture (FRIA-Belgium); the Agentschap voor Innovatie door Wetenschap en Technologie (IWT-Belgium); the F.R.S.-FNRS and FWO (Belgium) under the ``Excellence of Science -- EOS" -- be.h project n.\ 30820817; the Beijing Municipal Science \& Technology Commission, No. Z191100007219010 and Fundamental Research Funds for the Central Universities (China); the Ministry of Education, Youth and Sports (MEYS) of the Czech Republic; the Shota Rustaveli National Science Foundation, grant FR-22-985 (Georgia); the Deutsche Forschungsgemeinschaft (DFG), under Germany's Excellence Strategy -- EXC 2121 ``Quantum Universe" -- 390833306, and under project number 400140256 - GRK2497; the Hellenic Foundation for Research and Innovation (HFRI), Project Number 2288 (Greece); the Hungarian Academy of Sciences, the New National Excellence Program - \'UNKP, the NKFIH research grants K 131991, K 133046, K 138136, K 143460, K 143477, K 146913, K 146914, K 147048, 2020-2.2.1-ED-2021-00181, and TKP2021-NKTA-64 (Hungary); the Council of Science and Industrial Research, India; ICSC -- National Research Center for High Performance Computing, Big Data and Quantum Computing and FAIR -- Future Artificial Intelligence Research, funded by the NextGenerationEU program (Italy); the Latvian Council of Science; the Ministry of Education and Science, project no. 2022/WK/14, and the National Science Center, contracts Opus 2021/41/B/ST2/01369 and 2021/43/B/ST2/01552 (Poland); the Funda\c{c}\~ao para a Ci\^encia e a Tecnologia, grant CEECIND/01334/2018 (Portugal); the National Priorities Research Program by Qatar National Research Fund;  MCIN/AEI/10.13039/501100011033, ERDF ``a way of making Europe", and the Programa Estatal de Fomento de la Investigaci{\'o}n Cient{\'i}fica y T{\'e}cnica de Excelencia Mar\'{\i}a de Maeztu, grant MDM-2017-0765 and Programa Severo Ochoa del Principado de Asturias (Spain); the Chulalongkorn Academic into Its 2nd Century Project Advancement Project, and the National Science, Research and Innovation Fund via the Program Management Unit for Human Resources \& Institutional Development, Research and Innovation, grant B37G660013 (Thailand); the Kavli Foundation; the Nvidia Corporation; the SuperMicro Corporation; the Welch Foundation, contract C-1845; and the Weston Havens Foundation (USA).
\end{acknowledgments}

\bibliography{auto_generated}
\clearpage
\appendix
\numberwithin{figure}{section}
\section{Additional projections of the 2D fit\label{app}}

\begin{figure*}[ht]
	\centering
        \includegraphics[width=0.48\textwidth] {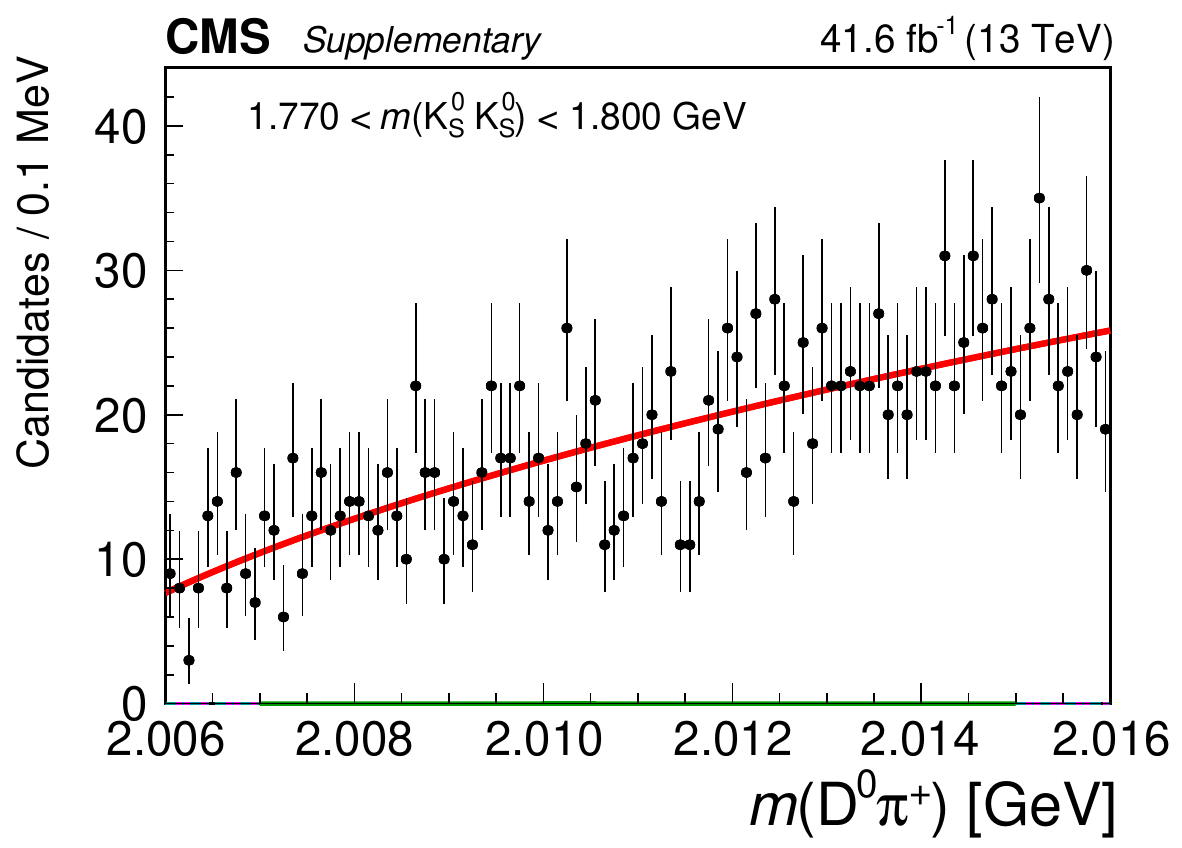}
        \includegraphics[width=0.48\textwidth] {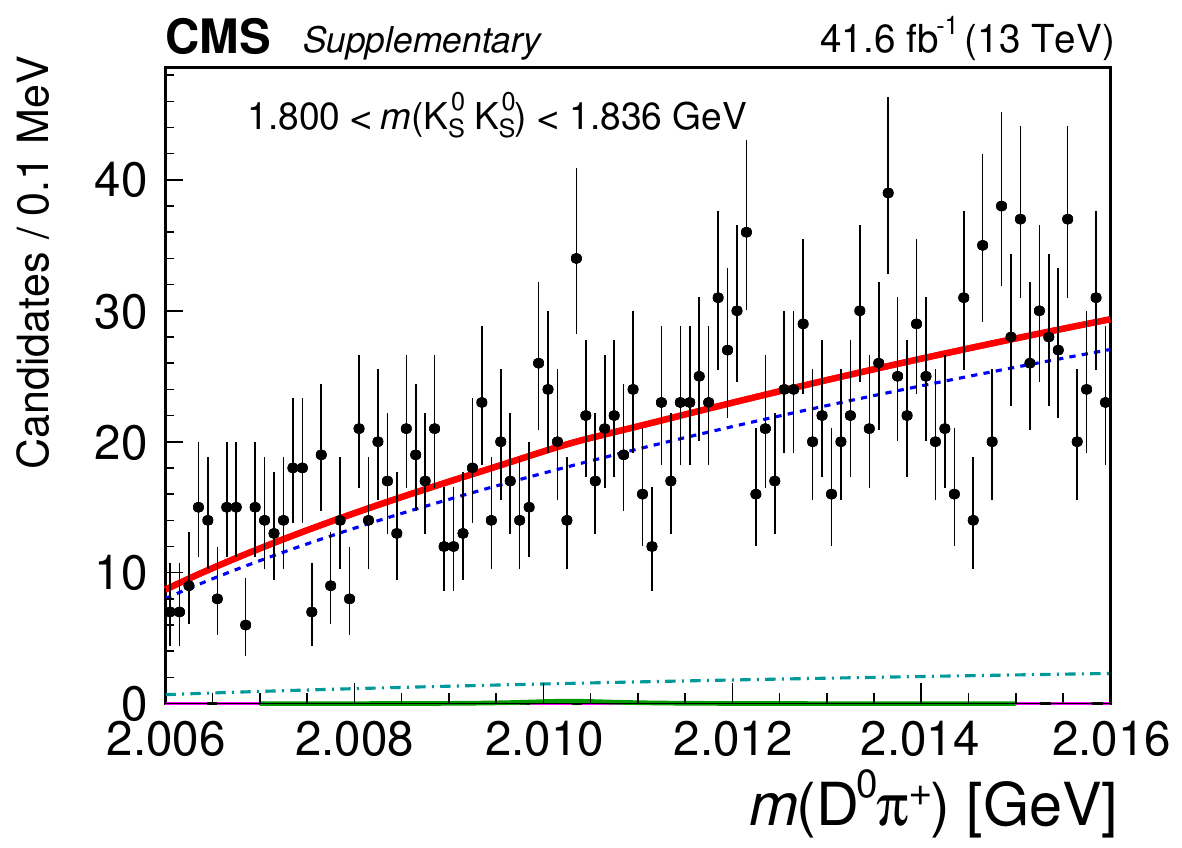}
        \includegraphics[width=0.48\textwidth] {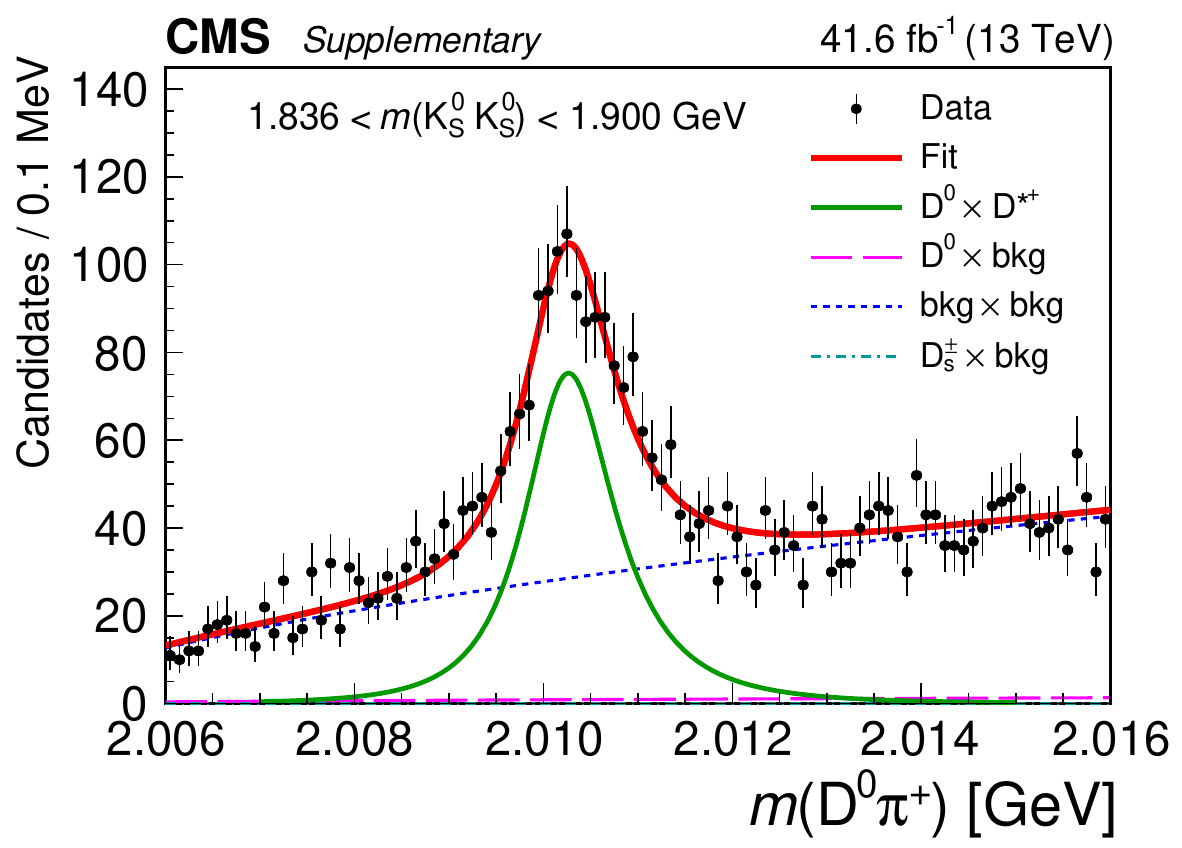}
        \includegraphics[width=0.48\textwidth] {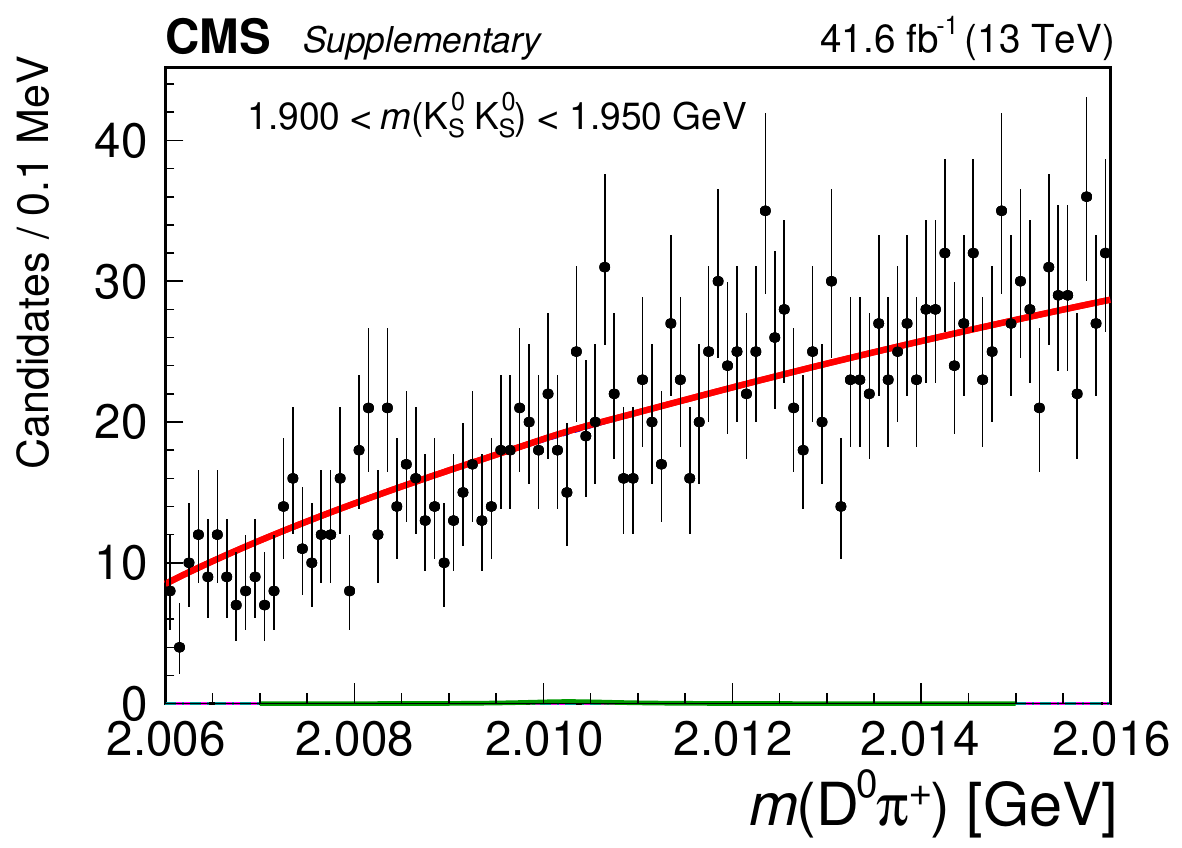}
        \includegraphics[width=0.325\textwidth]{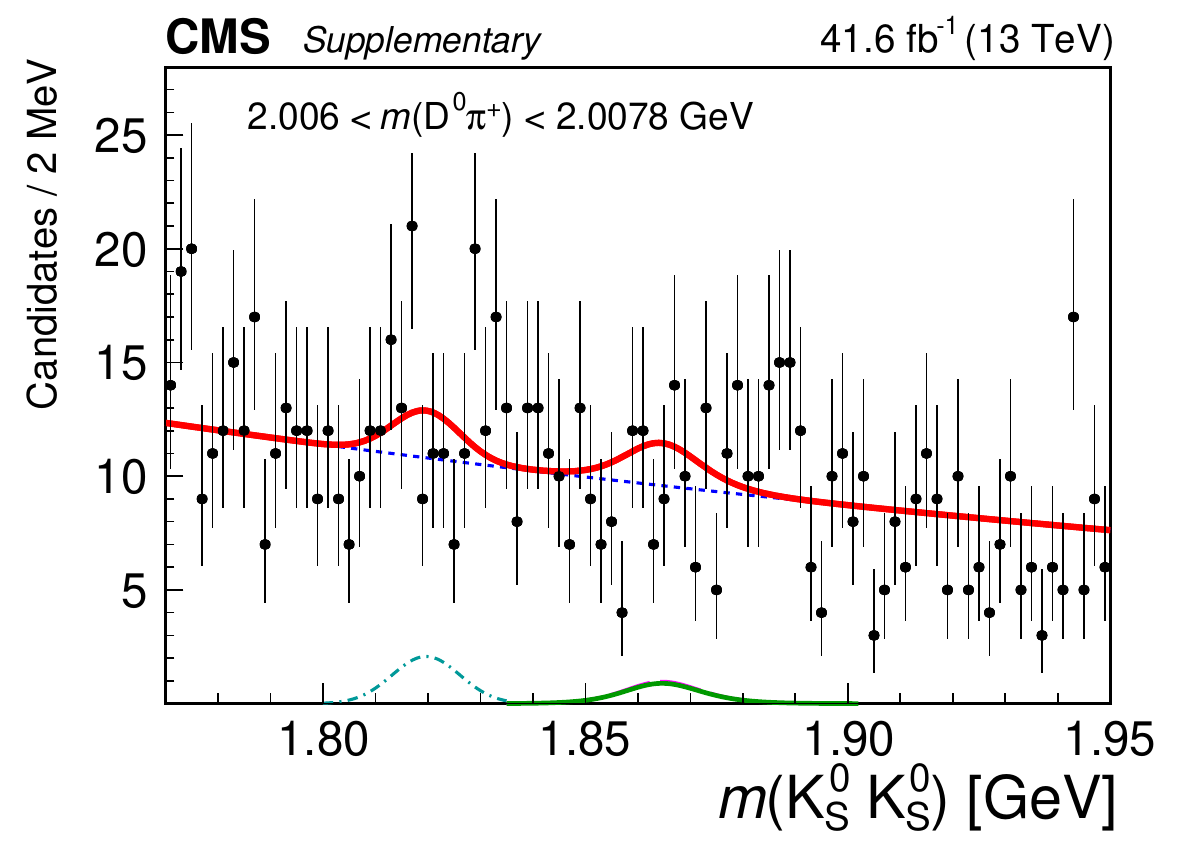}
        \includegraphics[width=0.325\textwidth]{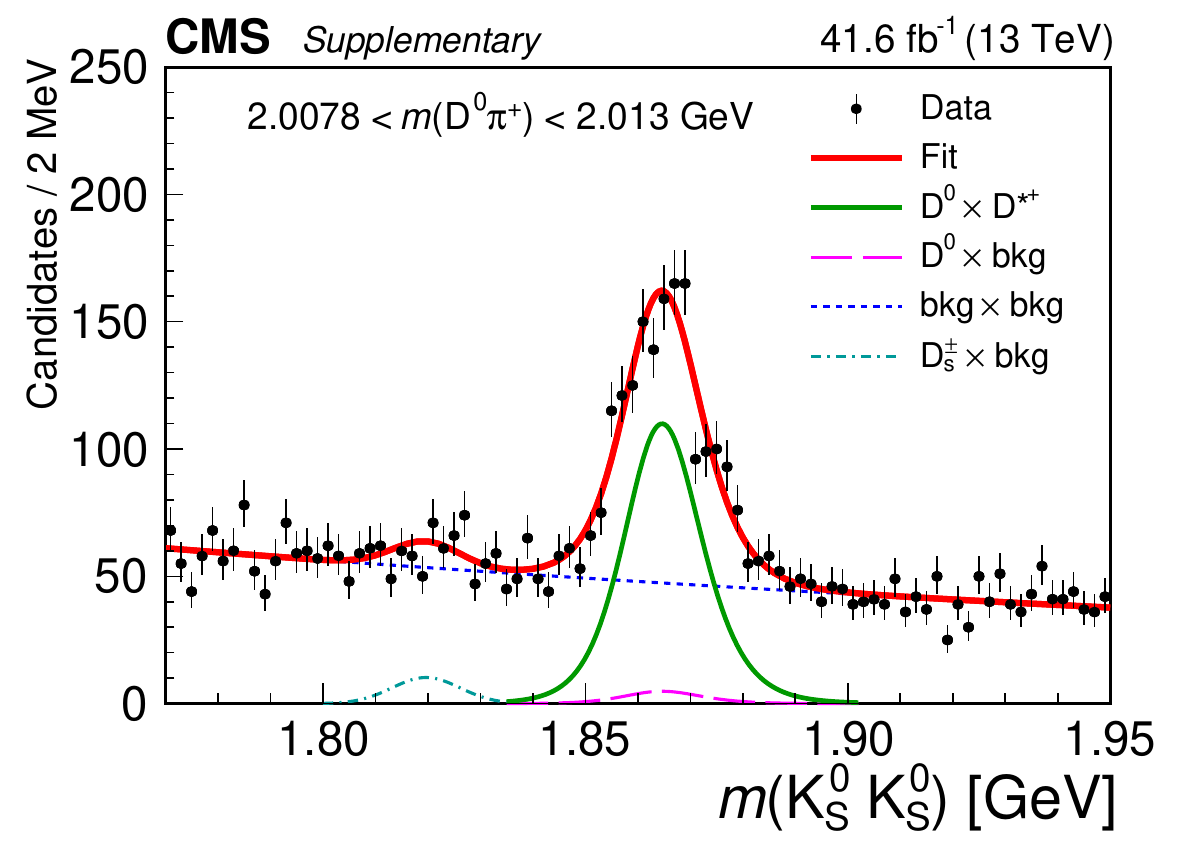}
        \includegraphics[width=0.325\textwidth]{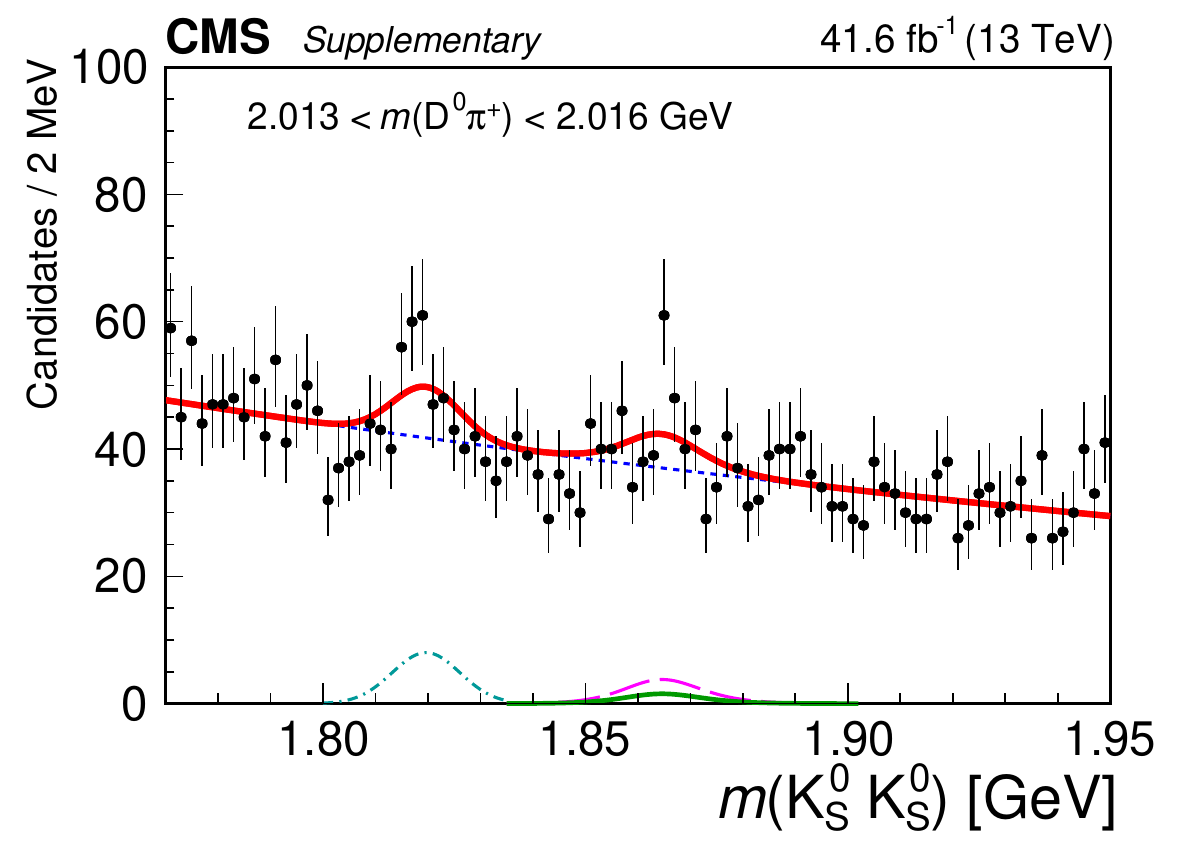}
	\caption{ Results of the 2D fit to the $\mDPi \times \mKSKS$ for the signal channel, \PDstp candidates.
		Upper and middle rows show 1D projections of the 2D fit on $m(\DzPip)$ in ranges of \mKSKS:
		left sideband (upper left), region of \DsKSKSPI contamination (upper right),
		signal region of \KSKS (middle left), and right sideband (middle right).
		Lower row shows 1D projections of the 2D fit on \mKSKS in ranges of $m(\DzPip)$: left sideband (left),
		signal region of \DzPip (center), and right sideband (right). 
		\label{figa:TwoDfitPplus}}
\end{figure*}

\begin{figure*}[ht]
	\centering
        \includegraphics[width=0.48\textwidth] {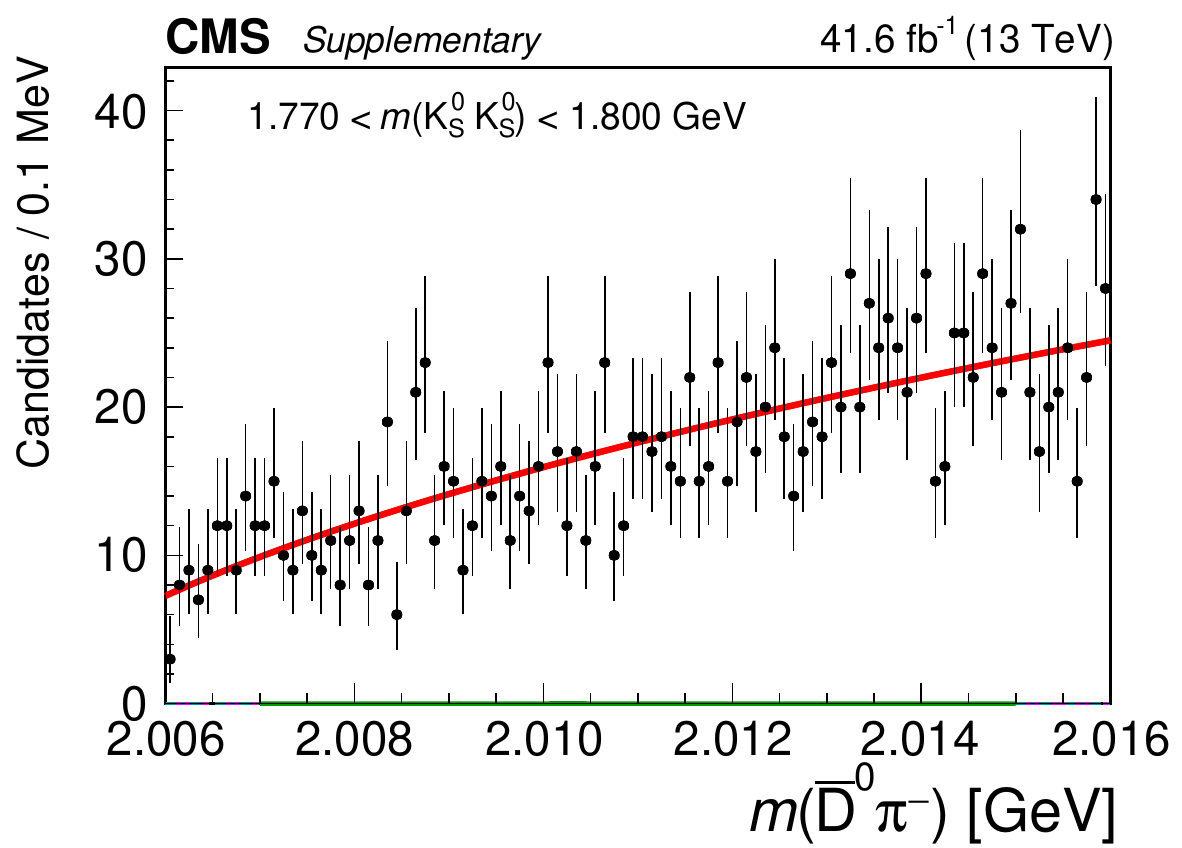}
        \includegraphics[width=0.48\textwidth] {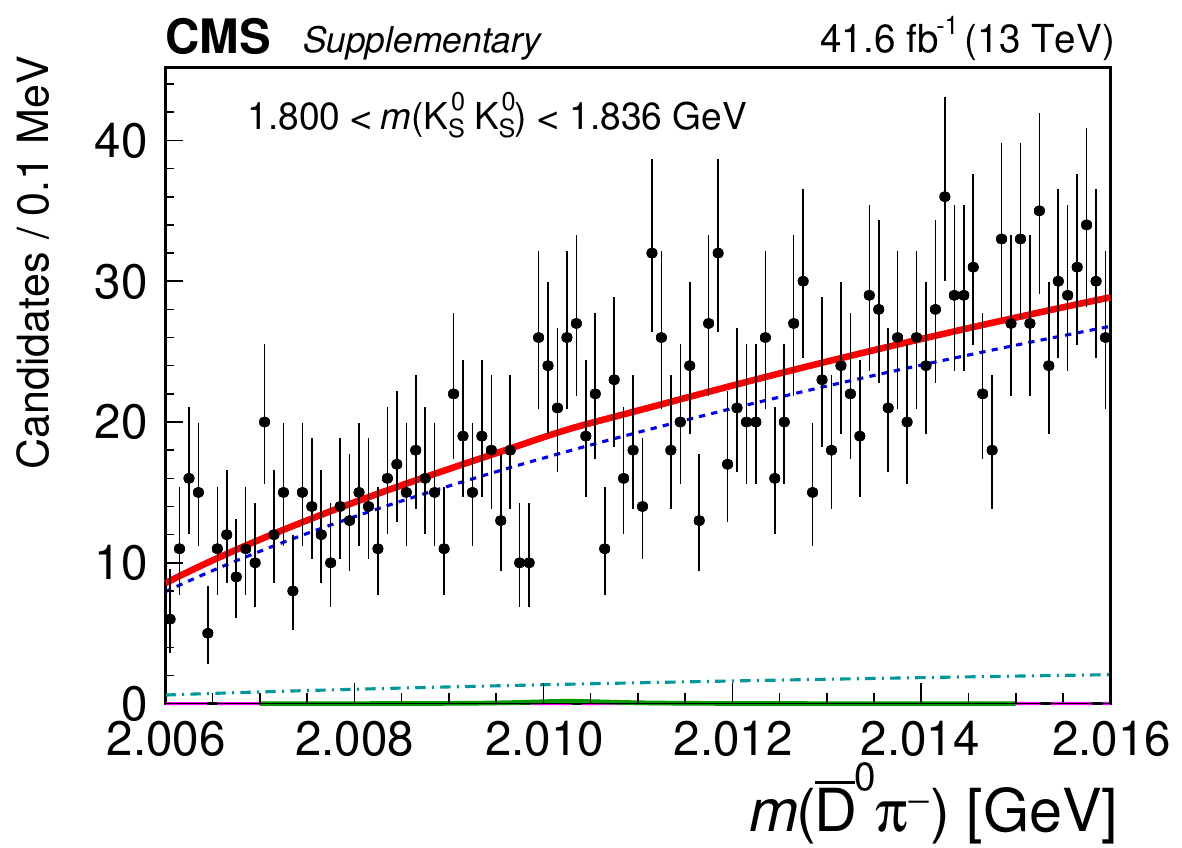}
        \includegraphics[width=0.48\textwidth] {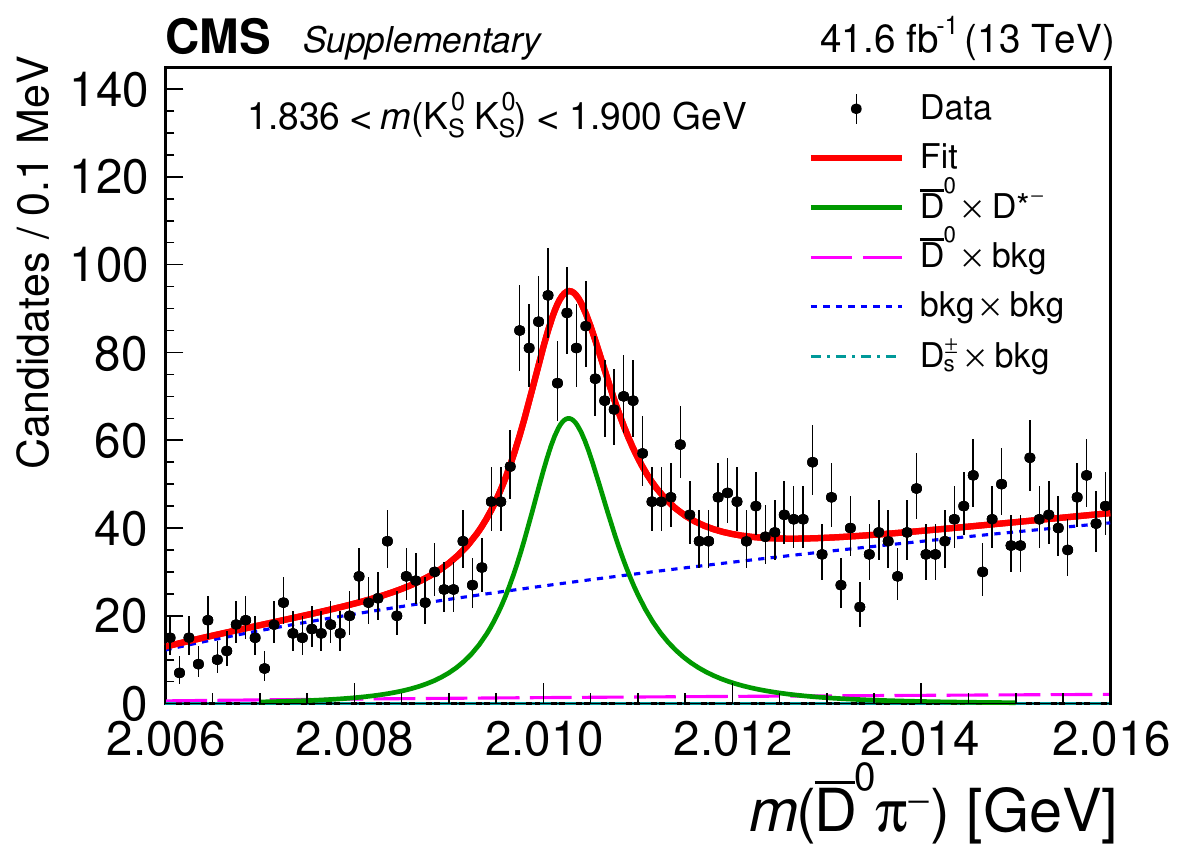}
        \includegraphics[width=0.48\textwidth] {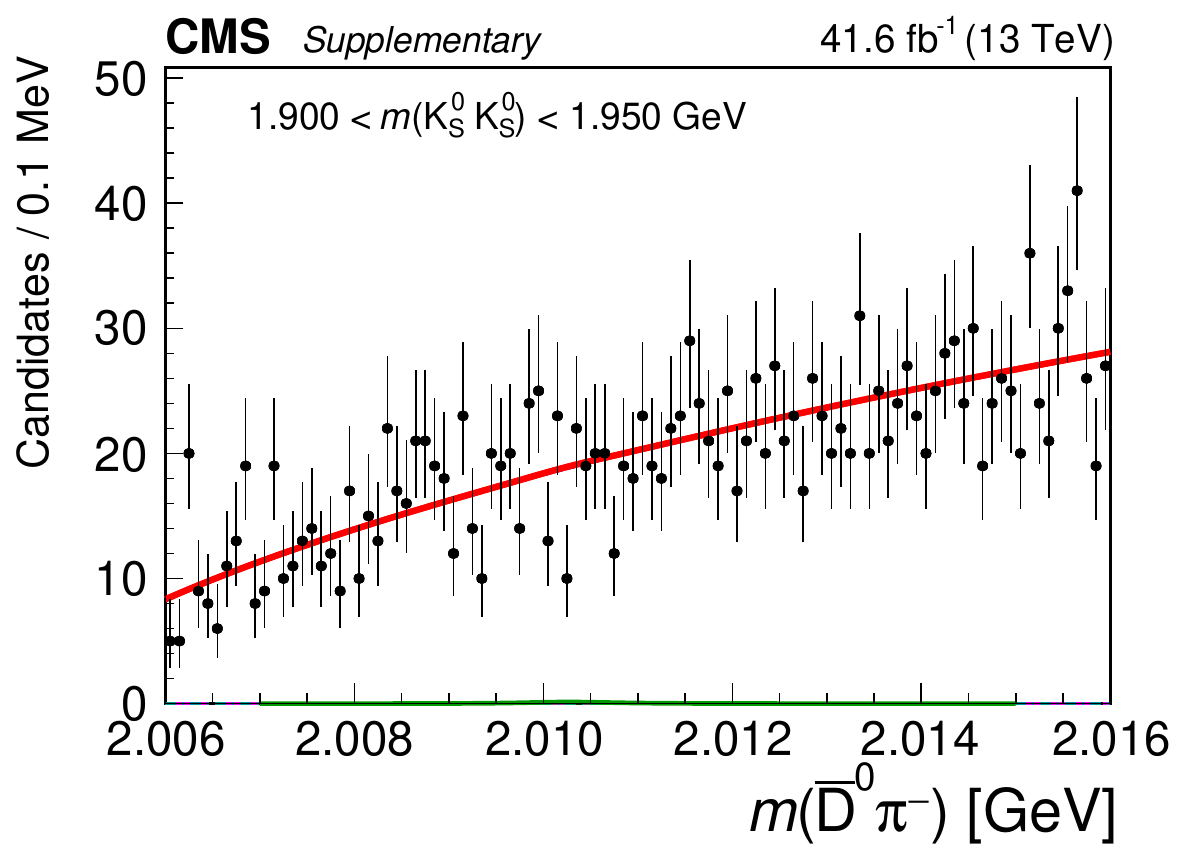}
        \includegraphics[width=0.325\textwidth]{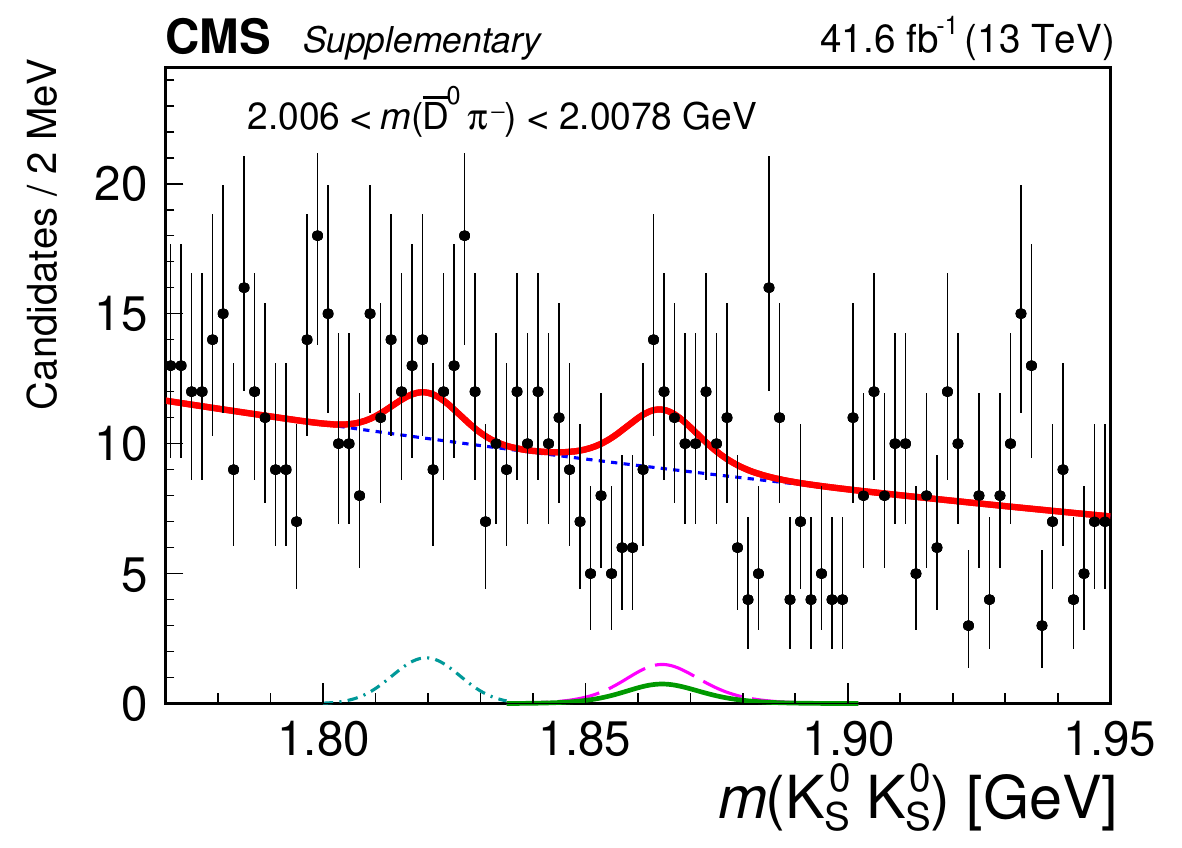}
        \includegraphics[width=0.325\textwidth]{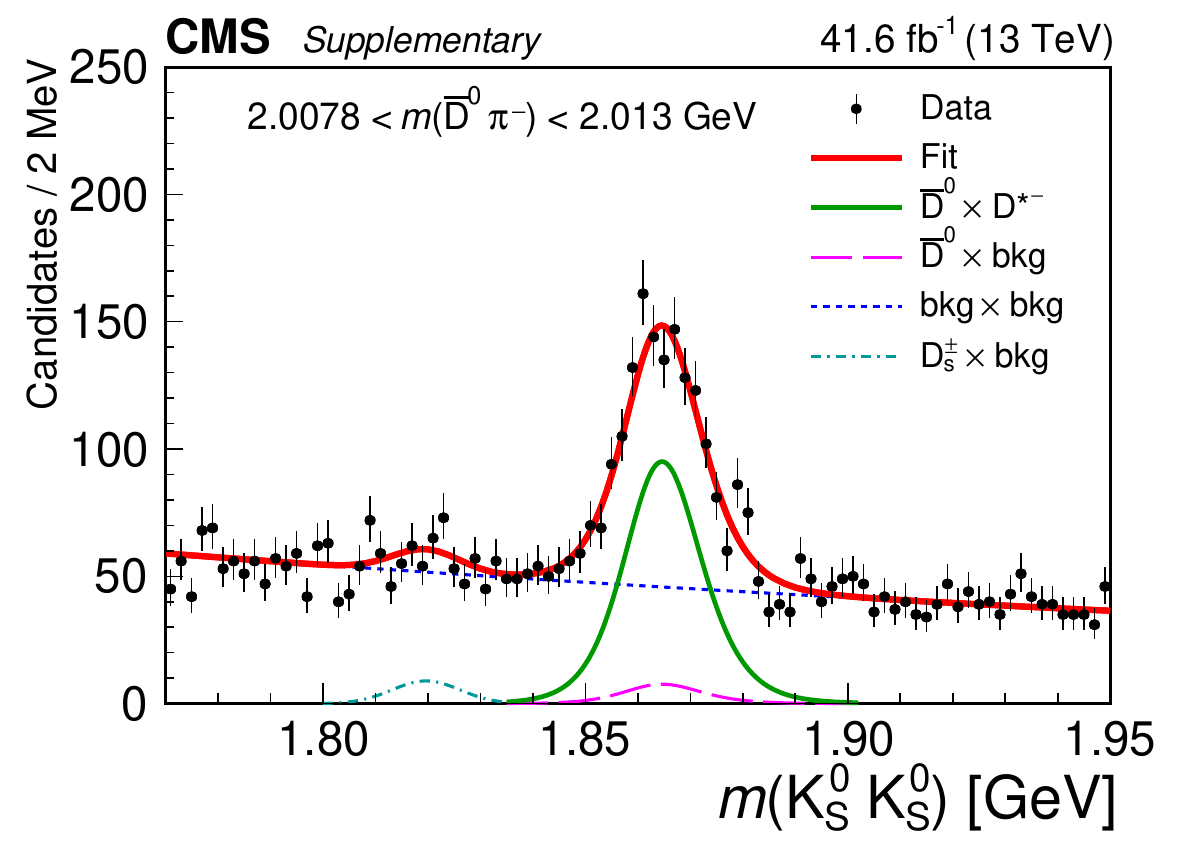}
        \includegraphics[width=0.325\textwidth]{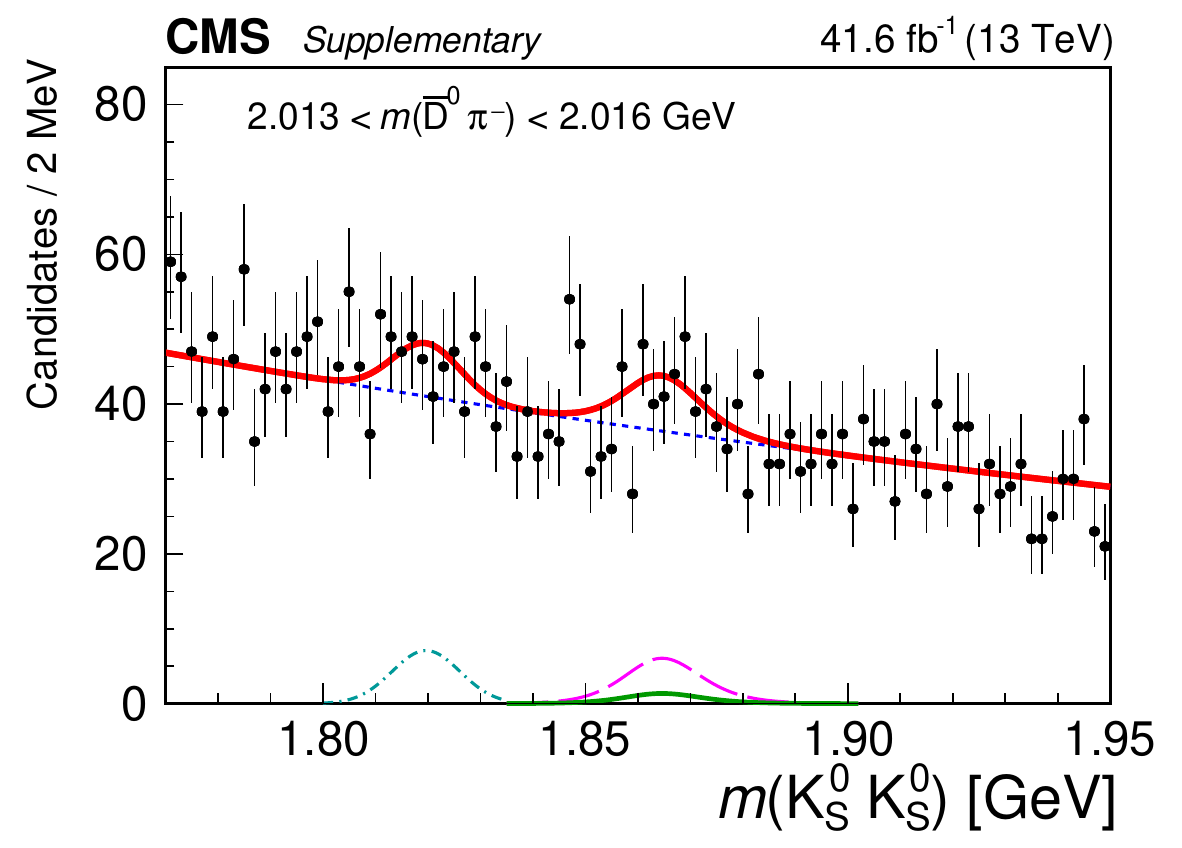}
	\caption{  Results of the 2D fit to the $\mDPi \times \mKSKS$ for the signal channel, \PDstm candidates.
		Upper and middle rows show 1D projections of the 2D fit on $m(\DzbPim)$ in ranges of \mKSKS:
		left sideband (upper left), region of \DsKSKSPI contamination (upper right),
		signal region of \KSKS (middle left), and right sideband (middle right).
		Lower row shows 1D projections of the 2D fit on \mKSKS in ranges of $m(\DzbPim)$: left sideband (left),
		signal region of \DzbPim (center), and right sideband (right).
		\label{figa:TwoDfitPminus}}
\end{figure*}

Figures~\ref{figa:TwoDfitPplus} and \ref{figa:TwoDfitPminus} 
show the projections of the 2D fit in the signal channel in subranges of the mass variables.
The top and middle rows show 1D projections of the 2D fit on \mDPi in ranges of \mKSKS:
left sideband, region of \DsKSKSPI contamination,
signal region of \KSKS, and right sideband.
The lower three plots show 1D projections of the 2D fit on \mKSKS in ranges of \mDPi: left sideband,
signal region of \DPi, and right sideband.

\cleardoublepage \section{The CMS Collaboration \label{app:collab}}\begin{sloppypar}\hyphenpenalty=5000\widowpenalty=500\clubpenalty=5000
\cmsinstitute{Yerevan Physics Institute, Yerevan, Armenia}
{\tolerance=6000
A.~Hayrapetyan, A.~Tumasyan\cmsAuthorMark{1}\cmsorcid{0009-0000-0684-6742}
\par}
\cmsinstitute{Institut f\"{u}r Hochenergiephysik, Vienna, Austria}
{\tolerance=6000
W.~Adam\cmsorcid{0000-0001-9099-4341}, J.W.~Andrejkovic, T.~Bergauer\cmsorcid{0000-0002-5786-0293}, S.~Chatterjee\cmsorcid{0000-0003-2660-0349}, K.~Damanakis\cmsorcid{0000-0001-5389-2872}, M.~Dragicevic\cmsorcid{0000-0003-1967-6783}, P.S.~Hussain\cmsorcid{0000-0002-4825-5278}, M.~Jeitler\cmsAuthorMark{2}\cmsorcid{0000-0002-5141-9560}, N.~Krammer\cmsorcid{0000-0002-0548-0985}, A.~Li\cmsorcid{0000-0002-4547-116X}, D.~Liko\cmsorcid{0000-0002-3380-473X}, I.~Mikulec\cmsorcid{0000-0003-0385-2746}, J.~Schieck\cmsAuthorMark{2}\cmsorcid{0000-0002-1058-8093}, R.~Sch\"{o}fbeck\cmsorcid{0000-0002-2332-8784}, D.~Schwarz\cmsorcid{0000-0002-3821-7331}, M.~Sonawane\cmsorcid{0000-0003-0510-7010}, S.~Templ\cmsorcid{0000-0003-3137-5692}, W.~Waltenberger\cmsorcid{0000-0002-6215-7228}, C.-E.~Wulz\cmsAuthorMark{2}\cmsorcid{0000-0001-9226-5812}
% ...
\par}
\vskip\cmsinstskip
\dag:~Deceased\\
$^{1}$Also at Yerevan State University, Yerevan, Armenia\\
$^{2}$Also at TU Wien, Vienna, Austria\\
% ...
\end{sloppypar}
\end{document}